\documentclass[
preprintnumbers,
twocolumn,
amsmath,amssymb,
 aps,
]{revtex4-2}

\usepackage{graphicx}
\usepackage{dcolumn}
\usepackage{bm}
\usepackage{hyperref}

\usepackage{comment}
\usepackage{caption}
\usepackage{subcaption}
\usepackage{booktabs}
\numberwithin{equation}{section}
\usepackage{color}

\usepackage{url}
\hypersetup{colorlinks=true, citecolor=cyan}

\newcommand{\be}{\begin{equation}}
\newcommand{\ee}{\end{equation}}

\usepackage{ulem}

\definecolor{mygreen}{RGB}{0,130,0} 

\begin{document}
\preprint{RESCEU-21/23}

\title{Physically consistent gravitational waveform for capturing\\ beyond general relativity effects in the compact object merger phase}
\author{Daiki Watarai${}^{1,2}$, Atsushi Nishizawa$^{3,2}$, and Kipp Cannon${}^{2}$}

\affiliation{${}^1$Graduate School of Science, The University of Tokyo, Tokyo 113-0033, Japan, \\${}^2$Research Center for the Early Universe (RESCEU), Graduate School of Science, The University of Tokyo, Tokyo 113-0033, Japan, \\${}^3$Physics Program, Graduate School of Advanced Science and Engineering, Hiroshima University, Higashi-Hiroshima, Hiroshima 739-8526, Japan}

\newcommand*{\diff}{\,\mathrm{d}}

\date{\today}

\begin{abstract}
The merger phase of compact binary coalescences is the strongest gravity regime that can be observed. To test the validity of general relativity (GR) in strong gravitational fields, we propose a gravitational waveform parameterized for deviations from GR in the dynamical and nonlinear regime of gravity. Our fundamental idea is that perturbative modifications to a GR waveform can capture possible deviations in the merger phase that are difficult to model in a specific theory of gravity. One of notable points is that our waveform is physically consistent in the sense that the additional radiative losses of energy and angular momentum associated with beyond-GR modifications are included. Our prescription to ensure physical consistency in the whole coalescence process is expected to be applicable to any deviation from the standard model of compact binary coalescence, such as the extended models of gravity or the environmental effects of compact objects, as long as perturbative modifications are considered. Based on the Fisher analysis and the compatibility with Einstein-dilaton Gauss-Bonnet waveforms, we show that our parameterization is a physically-consistent minimal one that captures the deviations in the nonlinear regime.

\end{abstract}

\maketitle


\section{\label{sec:level1}INTRODUCTION}
General relativity (GR), the standard theory of gravity, is consistent with experimental and observational tests to date, particularly in weak gravitational fields (for reviews, e.g.,~\cite{Will2014,Berti_2015}).
On the other hand, it is only recently that we have been able to verify GR in strong and dynamical gravity regimes. It has been made possible by the detection of gravitational waves (GWs) from compact binary coalescences (CBCs)~\cite{first_detection_gw150914, testing_GR_GW150914}.
In fact, the LIGO-Virgo-KAGRA (LVK) collaboration has so far reported 90 GW events from CBCs, and several tests have been performed~\cite{testing_GR_GWTC1,testing_GR_GWTC2, testing_GR_GWTC3} and found no significant evidence of GR breaking within the current accuracy.

From a theoretical point of view, GR is not considered as an ultimate theory of gravity. This is because it has several problems in the strong gravity limit, such as non-renormalizability as a quantum theory of gravity or the prediction of singularities where the laws of physics break down. Our understanding of physics in the strong gravitational fields is still poor due to a lack of observational confirmation. To address this issue with GW observations, it is crucial to extract information on the strong gravity regime from observational data. Since some theories predict modifications in GW signals  
(e.g.,~\cite{Berti2018,PhysRevD.94.084002}), analyzing GW data from CBCs allows us not only to confirm GR but might provide a clue on a theory beyond GR.

In general, a binary black hole (BBH) coalescence consists of three stages: inspiral, merger (or plunge), and ringdown phases~\cite{first_detection_gw150914}. Except for the merger phase, we can use the perturbative approaches, that is, the post-Newtonian expansion for the inspiral phase (e.g.,~\cite{Blanchet2014}) and the BH perturbation theory for the ringdown phase (e.g.,~\cite{Sasaki2003}). However, to understand the dynamics of the merger phase, we need to resort to numerical relativity (NR) simulations. This is because the merger is a highly dynamical gravity regime, and there is no analytical approach so far. Therefore, to estimate the source properties from the full inspiral-merger-ringdown (IMR) data, phenomenological waveforms are used, which are constructed by combining the merger waveform, which are derived by fitting to NR data, with analytical waveforms for the inspiral and ringdown parts.

In GW analysis for testing GR, parameterized frameworks are often used to quantify deviations from GR with clear physical interpretations (e.g.,~\cite{PPE_2009, Mishra_2010, Li_2012, Saleem_2022, Mehta_2023} for the inspiral and~\cite{Gosan_2012, Meidam_2014, Glampedakis_2017, Brito_2018, Cardoso_2019, McManus_2019, Carullo_2019, Ghosh_2021, Isi_2021} for the ringdown). One of the representatives is the parameterized post-Einsteinian framework~\cite{PPE_2009}. In \cite{PPE_2009}, Yunes and Pretorius introduced phenomenological modifications to the inspiral waveform, that can reproduce deviations for several theories that are calculated analytically. In contrast, for the merger, it is more difficult to construct a framework compatible with extended models due to few NR predictions based on specific models. This is primarily because the well-posedness of the initial value problems in the extended models is not known. Instead of performing the full NR simulations, several studies provide IMR waveforms for a limited number of extended theories in GR (e.g.,~\cite{dCS_waveform, EdGB_waveform}).

On the other hand, model-agnostic analyses for the merger part have been conducted so far. For example, LVK testing GR papers have shown the results of parameterized tests and IMR consistency tests~\cite{testing_GR_GW150914,testing_GR_GWTC1,testing_GR_GWTC2,testing_GR_GWTC3}. The parameterized tests estimate fractional deviation in one of GR coefficients, such as post-Newtonian parameters in the inspiral waveform or the ringdown and damping frequencies for the ringdown waveform. For the merger waveform, deviations of the artificial parameters introduced in the fitting procedure are estimated. One of the problems with the merger waveform is difficulty in the identification of the physical meanings of these fitting parameters. This means that, if a significant deviation is found by future analysis, it is difficult to interpret the deviation physically. Moreover, the analysis does not take into account the correlations between the parameters as only a single parameter is allowed to vary at a time. Such an analysis could bias the estimated results.
The IMR consistency tests check the consistency between the inspiral and the post-inspiral parts by assuming GR. This test simply checks consistency with GR, but it is difficult to derive the physical implications, because the post-inspiral part includes not only the merger part but also the late-inspiral and ringdown parts. This indicates that, if an inconsistency is found, it is difficult to derive a physical picture only for the nonlinear regime. Besides the LVK studies, there are a few studies that propose analytic frameworks and perform the analysis for the nonlinear region~\cite{testing_gr_Maggio-san, Bonnila-san}. However, a satisfactory method is still under discussion whether or not these waveform models can cover the IMR signal of a particular theory.

\subsection{Summary of our parameterized waveform}
To address the issues above, we propose a physically consistent modified IMR waveform model that can measure perturbative deviations from GR in the nonlinear regime by introducing two beyond-GR parameters: one each to phase and amplitude. Our modification is minimal in the sense that amplitude and phase can be modified independently. In this work, we take the IMRPhenomD waveform \cite{IMRPhenomD_2} as our fiducial waveform and do not modify the inspiral and ringdown parts since we focus on the deviations in the merger part.

\begin{itemize}
\item{\textbf{Beyond-GR phase parameter}}\\
We adopt the largest principal component among the artificial parameters in the phase of a phenomenological GR waveform. We believe that the principal component contains physical information about the nonlinear dynamics of BBH, including orbital evolution and energy and angular momentum loss rates, since the PCA is considered to be the most dominant independent component of the merger part, although it is not easy to extract physical information about the PCA itself. Furthermore, this specification is expected to break degeneracies between artificially introduced parameters, which have not been considered in previous studies
(e.g.,~\cite{testing_GR_GW150914,testing_GR_GWTC1,testing_GR_GWTC2}). 
\item{\textbf{Beyond-GR amplitude parameter}}\\
We adopt a parameter that describes an amplification after the inspiral phase. This parameter is basically given as the parameter that characterizes the amplitude peak of a signal.
\item{\textbf{Additional radiation backreaction}}\\
To ensure physical consistency for the entire IMR process, we include the radiation reaction to our waveform by calculating the additional losses of energy and angular momentum associated with the beyond-GR amplitude parameter. These losses are calculated based on the quadrupole formulas for energy and angular momentum losses and then are reflected in the mass and spin of a remnant BH, resulting in a physically consistent waveform. 
\end{itemize}
Our prescription to ensure the physical consistency is expected to be applicable to any deviation from the standard model of compact binary coalescence, such as the extended models of gravity or the environmental effects of BHs, as long as perturbative modifications are considered. Our waveform captures deviations only in the nonlinear regime and, importantly, is constructed so as to ensure the physical consistency of a remnant BH. Finally, we show that our waveform can cover Einstein-dilaton Gauss-Bonnet gravity waveforms~\cite{EdGB_waveform} within the measurement errors in the O4 and O5 observations. 

\subsection{Plan of the paper}
The rest of this paper is organized as follows. In Sec.~\ref{sec:GR_waveform}, details of IMRPhenomD waveform \cite{IMRPhenomD_2}, which is adopted as a basis for our modified waveform, are presented. In Sec.~\ref{sec:Modified_waveform}, the construction of our modified waveform is discussed. We elaborate on the strategies for introducing beyond-GR parameters into phase and amplitude, and for ensuring physical consistency through the inclusion of radiation backreaction. In Sec.~\ref{sec:Systematic_studies}, we evaluate the properties of our waveform and the measurability of beyond-GR parameters. Specifically, the mismatch, the measurement error estimates in the O4 and O5 observations, and the compatibility with Einstein dilaton Gauss-Bonnet gravity waveforms~\cite{EdGB_waveform} are presented. Finally, in Sec.~\ref{sec:Discussion} and \ref{sec:Conclusion}, the discussions and conclusions are presented, respectively. Throughout this paper, we adopt the geometrical unit where $G=c=1$ except for Sec.~\ref{sec:physical_consistency}.

\section{GR waveform}
\label{sec:GR_waveform}
Extracting the physical information imprinted in the GWs requires waveform templates, which are a family of the waveforms that are parameterized by many binary parameters and allow us to estimate a set of the best-fitting parameters. In the inspiral and ringdown, we can construct analytical waveforms because perturbative approaches can be applied. For the inspiral, we use the post-Newtonian (PN) approximation method (e.g.,~\cite{Blanchet2014}), which gives an accurate expression for a waveform when binary separation is large and slow velocity, $v \ll c$. For the ringdown, the waveform can be modeled based on BH perturbation theory (e.g.,~\cite{Sasaki2003}), which is valid as long as deviation from an unperturbed BH spacetime can be considered small. However, for the merger phase we need full numerical relativity (NR) simulations despite high computational cost. To avoid this, numerical fitting to the NR waveforms~\cite{PhysRevD.79.129901} are used. The waveforms are approximate analytic ones and are easy to handle in GW searches more cheaply and quickly.
Several Phenom waveforms are currently available for practical use. In this study, we adopt IMRPhenomD waveform~\cite{IMRPhenomD_2}, which is one of the Phenom waveforms for an aligned-spinning binary with circular orbit.

In this section, we briefly review 
the IMRPhenomD model, which describes the dominant GW mode from the spin-aligned BBH coalescence process. 

\subsection{Conventions}
We begin by confirming the convention used throughout this paper. A BBH system originally has 8 intrinsic parameters: BH masses, $m_i$, and spin vectors, $\Vec{S}_i$, for each component $(i=1,2)$. We define the $z$ direction as the direction of the orbital angular momentum and use some specific combinations, 
\begin{gather}
    M := m_1 + m_2 \:\::\mathrm{total\:mass}\;,\label{totall_mass}\\
    S := (\Vec{S}_1 + \Vec{S}_2)\cdot\hat{L}: \:\:
    \begin{split}
    &\mathrm{spin\:component\:along}\\
    &\mathrm{orbital\:angular\:momentum}    
    \end{split}\;, \label{total_spin}\\
    \chi_i := \frac{\Vec{S}_i\cdot\hat{L}}{{m_i}^2}\:\::\mathrm{dimensionless\:spin\:parameter}\label{chi_i}\;,
\end{gather}
where $\hat{L}$ is the unit vector directed to the vector of orbital angular momentum.
Furthermore, we define other parameters,
\begin{gather}
    \eta := \frac{m_1 m_2}{M^2}\:\::\mathrm{symmetric\:mass\:ratio}\label{symmetric_mass_ratio}\;,\\
    \hat{S} := \frac{S}{1-2\eta} \in [-1,1]\:\::\mathrm{normalized\:}S \label{normalized_total_spin}\;,\\
    \chi_\mathrm{eff} := \frac{m_1\chi_1+m_2\chi_2}{M}\:\::\mathrm{effective\:spin\:parameter}\label{chi_eff}\;,\\
    \chi_{\mathrm{PN}} := \chi_\mathrm{eff} - \frac{38\eta}{113}(\chi_1+\chi_2)\: \\ :\mathrm{leading\:spin\:effect\:in\:PN\:expansion}\;\label{chi_PN}.
\end{gather}
In this work, we restrict ourselves to spin-aligned binaries. Thus, there are four intrinsic parameters, the mass, $m_i$, and the component spin, $S_i := \Vec{S_i}\cdot\hat{L}$.

In the paper, we define the Fourier transform of $h(t)$ as 
\begin{equation}
    \tilde{h}(f) = \int^{\infty}_{-\infty} \diff t \;h(t)\mathrm{e}^{-2\pi i ft}\;. 
\end{equation}
Furthermore, we adopt a definition of the noise-weighted inner product between two arbitrary functions, $A(f)$ and $B(f)$,
\begin{equation}
\label{inner_product}
    \left(A(f), B(f)\right) := 4\mathrm{Re}\left[\int^{f_\mathrm{max}}_{f_\mathrm{min}} \frac{A^\ast(f) B(f)}{S_n(f)}df\right]\;,
\end{equation}
where $f_\mathrm{min}$, $f_\mathrm{max}$ are the lower and upper cutoff frequencies and $S_n(f)$ is the noise power spectral density of a detector.

\subsection{IMRPhenomD}

The IMRPhenomD waveform describes the dominant $(l,m)=(2,\pm 2)$ mode from the entire process of an aligned-spin (non-precessing) BBH merger in the frequency domain, with the mass ratio up to $1:18$ \cite{IMRPhenomD_2}. The form is written as
\begin{equation}
\label{IMRPhenomD}
    \tilde{h}_\mathrm{GR}(f) = A_\mathrm{GR}(f)\:\mathrm{e}^{-i\phi_\mathrm{GR}(f)}\;,
\end{equation}
where $\phi_\mathrm{GR}(f)$ and $A_\mathrm{GR}(f)$ are the phase and amplitude of a GW signal, respectively. In the following, we fix a subscript, GR, to a quantity in GR when we emphasize it. $\phi_\mathrm{GR}(f)$ and $A_\mathrm{GR}(f)$ are divided into three frequency ranges, denoted as the inspiral (ins), intermediate (int), and merger-ringdown (MR), respectively,
\begin{gather}
\label{PhenomD_phi}
\phi_\mathrm{GR}(f) = \left\{
\begin{array}{lll}
\phi_{\mathrm{ins}}(f), & f \leq f_{\mathrm{p1}}\\
\phi_{\mathrm{int}}(f), & f_{\mathrm{p1}} \leq f \leq f_{\mathrm{p2}}\\
\phi_{\mathrm{MR}}(f), & f \geq f_{\mathrm{p2}}
\end{array}\;,
\right.\\
\label{PhenomD_A}
A_\mathrm{GR}(f) = \left\{
\begin{array}{lll}
A_{\mathrm{ins}}(f), & f \leq f_{\mathrm{a1}}\\
A_{\mathrm{int}}(f), & f_{\mathrm{a1}} \leq f \leq f_{\mathrm{a2}}\\
A_{\mathrm{MR}}(f), & f \geq f_{\mathrm{a2}} 
\end{array}\;,
\right.
\end{gather}
where $f_{\mathrm{p}i}$ and $f_{\mathrm{a}i}$ ($i=1,2$) are the collocation frequencies introduced below.

\subsubsection{\textbf{Phenomenological parameters}}
To model a nonlinear regime, the IMRPhenomD waveform adopts phenomenological parameters, that are introduced to fit to NR waveforms.
$\phi_\mathrm{GR}(f)$ and $A_\mathrm{GR}(f)$ have 11 and 14 phenomenological parameters, respectively~\cite{IMRPhenomD_2}. Here we denote them all as $\{\Lambda^i\}$. 17 coefficients out of $\{\Lambda^i\}$ are modeled with polynomial functions of $\eta$ and $\chi_{\mathrm{PN}}$ from fitting to NR data,
\begin{equation}
\label{artificial_coefficients}
\begin{split}
    \Lambda^i(\eta,\chi_\mathrm{PN}) =  
    &\lambda^i_{00} + \lambda^i_{10}\eta \\
    &+ (\chi_{\mathrm{PN}}-1)(\lambda^i_{01} + \lambda^i_{11}\eta + \lambda^i_{21}\eta^2)\\
    &+(\chi_{\mathrm{PN}}-1)^2(\lambda^i_{02} + \lambda^i_{12}\eta + \lambda^i_{22}\eta^2)\\
    &+ (\chi_{\mathrm{PN}}-1)^3(\lambda^i_{03} + \lambda^i_{13}\eta + \lambda^i_{23}\eta^2)\:,
\end{split}
\end{equation}
where $\{\lambda^i_{jk}\}$ are given in \cite{IMRPhenomD_2}.

The remaining 8 coefficients are determined by the $C^1$ continuity conditions for $\phi_\mathrm{GR}(f)$ and $A_\mathrm{GR}(f)$ at the collocation frequencies, 
\begin{gather}
    f_{\mathrm{p1}} = 0.014 / M \;, \\
    f_{\mathrm{p2}} = 0.5 f_\mathrm{RD} \;,\label{col_p2} \\
    f_{\mathrm{a1}} = 0.018 / M \;, \\
    f_{\mathrm{a2}} = f_{\mathrm{peak}} \:,\label{col_a2} 
\end{gather}
with
\begin{equation}
    f_\mathrm{peak} = \left| f_\mathrm{RD} - \frac{\gamma_3 f_{\mathrm{damp}}\left(1-\sqrt{1-{\gamma_2}^2}\right)}{\gamma_2} \right|\:,
\end{equation}
where $\{\gamma_2, \gamma_3\}$ are the amplitude coefficients in the merger-ringdown regime determined by Eq.~(\ref{artificial_coefficients}), and $f_\mathrm{RD}$ and $f_\mathrm{damp}$ are the ringdown and damping frequencies, that are computed by the following fitting formulas~\cite{Berti_2009}:
\begin{gather}
    f_\mathrm{RD} = \frac{1}{2\pi M}\left\{1.5251 - 1.1568(1-a_\mathrm{f})^{0.1292}\right\} \;,\label{formula_f_RD} \\
    Q=0.7000 + 1.4187(1-a_\mathrm{f})^{-0.4990}\;,
    \label{formula_Q}\\
    f_\mathrm{damp} = \frac{f_\mathrm{RD}}{2Q}\:. \label{formula_f_damp}
\end{gather}
Here $a_\mathrm{f}$ is the dimensionless spin parameter of the remnant BH, which is defined as
\begin{equation}
\label{a_f}
    a_\mathrm{f} := \frac{cS_\mathrm{f}}{GM^2_\mathrm{f}}\;,
\end{equation}
where $M_\mathrm{f}$ and $ S_\mathrm{f}$ are mass and spin of the remnant, respectively.
For our study, we use the latest remnant formula of $a_\mathrm{f}$ given in~\cite{PhysRevD.95.064024}, while the original waveform adopts the formula given in~\cite{IMRPhenomD_1}.

In the next subsection, we see details of IMRPhenomD construction. Following \cite{IMRPhenomD_2}, we omit the total mass dependence. If you want to revive the dependence, replace $f$ with $Mf$. 

\subsubsection{\textbf{Phase}}
\label{sec:PhenomD_phase}
In the inspiral regime, $f \leq f_\mathrm{p1}$, phase, $\phi_{\mathrm{ins}}(f)$, is modeled based on the TaylorF2 waveform~\cite{Damour_2001, Damour_2002, Arun_2005}, which provides an analytical expression for the dominant GW modes, $(l,m=2, \pm 2)$, in the early inspiral,
\begin{gather}
    \tilde{h}_{\mathrm{TF2}}(f) = \sqrt{\frac{5\eta}{24}}\pi^{-2/3}\frac{M^2}{r} f^{-7/6}\mathrm{e}^{-i\phi_{\mathrm{TF2}}(f)}\;,\\
\begin{split}
    \phi_{\mathrm{TF2}}(f) = & 2\pi f t_c - \phi_c - \pi/4\\ &+ \frac{3}{128\eta} (\pi f)^{-5/3}\sum^{7}_{i=0}\phi_i(\pi f)^{i/3}\:,
\end{split}
\end{gather}
where $r$ is distance to a binary, $t_c$ and $\phi_c$ are the time and phase at coalescence, respectively, and $\{\phi_i\}$ are analytically calculated PN coefficients up to 3.5~PN order~\cite{Arun_2005}. 
Then, the phase part of the IMRPhenomD consists of the TaylorF2 phase and phenomenological corrections modeled based on the PN expansion,
\begin{equation}
\begin{split}
    \phi_{\mathrm{ins}}(f) = &\phi_{\mathrm{TF2}}(f)\\ &+ \frac{1}{\eta}\Big( \sigma_0 + \sigma_1 f + \frac{3}{4}\sigma_2 f^{4/3} + \frac{3}{5}\sigma_3 f^{5/3} + \frac{1}{2} \sigma_4 f^2 \Big)\:,
\end{split}
\end{equation}
where $\{\sigma_i\}$ are 4 fitting coefficients in the inspiral given by Eq.~(\ref{artificial_coefficients}).  

In the intermediate regime, $f_\mathrm{p1} \leq f \leq f_\mathrm{p2} $, the phase $\phi_{\mathrm{int}}(f)$ is given as
\begin{equation}
\label{phi_int}
    \phi_{\mathrm{int}}(f) = \frac{1}{\eta}\left(\beta_0 + \beta_1 f + \beta_2 \log{f} - \frac{\beta_3}{3} f^{-3} \right)\:,
\end{equation}
where $\{\beta_2, \beta_3\}$ are artificial coefficients in the intermediate regime given by Eq.~(\ref{artificial_coefficients}). 

In the merger-ringdown regime, $f \geq f_\mathrm{p2}$, the phase part is given as
\begin{equation}
\label{MR_phase}
\begin{split}
    \phi_{\mathrm{MR}}(f)=\frac{1}{\eta} \biggl\{ &\alpha_0 + \alpha_1 f - \alpha_2 f^{-1}\\ &+ \frac{4}{3} \alpha_3 f^{3/4} + \alpha_4\tan^{-1}{ \biggl( \frac{f-\alpha_5 f_\mathrm{RD}}{f_\mathrm{damp}}} \biggr) \biggr\}\:, \\   
\end{split}
\end{equation}
where $\{\alpha_2,\alpha_3,\alpha_4,\alpha_5\}$ are artificial coefficients in the merger-ringdown regime given by Eq.
(\ref{artificial_coefficients}). 
In particular, $\{\alpha_4,\alpha_5\}$ characterize the ringdown phase in the sense that arctan part, $\alpha_4\tan^{-1}{ [ ({f-\alpha_5 f_\mathrm{RD}})/{f_\mathrm{damp}]}}$, models an analytic behavior of the phase of the ringdown part. On the other hand, $\{\alpha_2, \alpha_3\}$, characterize the nonlinear region around merger. The remaining 4 coefficients, $\{\beta_0,\beta_1,\alpha_0,\alpha_1\}$, are fixed by imposing the $C^1$ continuity conditions at $f=f_\mathrm{p1}$ and $f_\mathrm{p2}$, that is,
\begin{gather}
    \phi_\mathrm{ins}(f_\mathrm{p1}) = \phi_\mathrm{int}(f_\mathrm{p1})\;, \\
    \phi_\mathrm{int}(f_\mathrm{p2}) = \phi_\mathrm{MR}(f_\mathrm{p2})\;, \\ 
    \phi'_\mathrm{ins}(f_\mathrm{p1}) = \phi'_\mathrm{int}(f_\mathrm{p1})\;, \\
    \phi'_\mathrm{int}(f_\mathrm{p2}) = \phi'_\mathrm{MR}(f_\mathrm{p2})\:,
\end{gather}
where the prime is a derivative with respect to $f$.

We can write these continuity conditions in the matrix form for $\{\beta_{0,1},\alpha_{0,1}\}$,
\begin{equation}
\label{phase_connection}
\begin{pmatrix}
1 & f_\mathrm{p1} & 0 & 0\\
0 & 1 & 0 & 0\\
1 & f_\mathrm{p2} & -1 & - f_\mathrm{p2}\\
0 & 1 & 0 & -1 \\
\end{pmatrix}
\begin{pmatrix}
\beta_0 \\ \beta_1 \\ \alpha_0 \\ \alpha_1
\end{pmatrix}
= 
\begin{pmatrix}
C_1 \\ C_2 \\ C_3 \\ C_4\\
\end{pmatrix}\:,
\end{equation}
with
\begin{gather}
    C_1 = \eta \phi_\mathrm{ins}(f_\mathrm{p1}) - \beta_2 \log{f_\mathrm{p1}} + \frac{\beta_3}{3} {f_\mathrm{p1}}^{-3}\;, \label{C1} \\   
    C_2 = \eta \phi'_\mathrm{ins}(f_\mathrm{p1}) - \beta_2 {f_\mathrm{p1}}^{-1} - \beta_3 {f_\mathrm{p1}}^{-3}\;,\label{C2}  \\  
    \label{C3}
    \begin{split}
    C_3 = & - \beta_2 \log{f_\mathrm{p2}} + \frac{\beta_3}{3}{f_\mathrm{p2}}^{-3}-\alpha_2 {f_\mathrm{p2}}^{-1} + \frac{4}{3}\alpha_3 {f_\mathrm{p2}}^{3/4} \\ &+ \alpha_4\tan^{-1}{ \biggl( \frac{f_\mathrm{p2}-\alpha_5 f_\mathrm{RD}}{f_\mathrm{damp}}}\biggr)\;, \\ 
    \end{split}\\
    \begin{split}
    C_4 = &- \beta_2 {f_\mathrm{p2}}^{-1} - \beta_3 {f_\mathrm{p2}}^{-4} + \alpha_2 {f_\mathrm{p2}}^{-2} + \alpha_3 {f_\mathrm{p2}}^{-1/4} \\ &+ \frac{\alpha_4 f_\mathrm{damp}}{{f_\mathrm{damp}}^2+(f_\mathrm{p2}-\alpha_5 f_\mathrm{RD})^2}\:. \label{C4}
    \end{split}
\end{gather}
$\{ C_i \}$ are constant values once we determined the initial configuration and the coefficients, $\{\beta_2, \beta_3\}$ and $\{\alpha_2,\alpha_3,\alpha_4,\alpha_5\}$.

\subsubsection{\textbf{Amplitude}}
\label{sec:PhenomD_amplitude}
In the inspiral regime, amplitude consists of the TaylorF2 part, $A_{\mathrm{PN}}(f)$ (up to 3\,PN order), and the phenomenological corrections above 3.5\,PN order,
\begin{equation}
    A_{\mathrm{ins}}(f) = A_{\mathrm{PN}}(f) + A_0(f) \sum^{6}_{i=0} \rho_i f^{(i+6)/3} \:,   
\end{equation}
with
\begin{equation}
    A_{\mathrm{PN}}(f) = A_0(f) \sum^6_{i=0} \mathcal{A}_i (\pi f)^{i/3}\:,  
\end{equation}
where $A_0(f):=\sqrt{\frac{2\eta}{3\pi}}f^{-7/6}$ is a normalized TaylorF2 amplitude in the sense that $f^{7/6}A_0(f)$ approaches unity at the limit of $f \rightarrow 0$. $\{\mathcal{A}_i\}$ are the PN coefficients derived analytically~\cite{IMRPhenomD_2}, and $\{\rho_i\}$ are 7 artificial coefficients in the inspiral regime given by Eq.~(\ref{artificial_coefficients}).

In the merger-ringdown regime, $f \geq f_\mathrm{a2}$, the amplitude is
\begin{equation}
    A_{\mathrm{MR}}(f) = A_0(f) \gamma_1 \frac{\gamma_3 f_{\mathrm{damp}}}{(f-f_{\mathrm{RD}})^2+ {\gamma_3}^2 {f_{\mathrm{damp}}^2}  } \mathrm{e}^{-\frac{\gamma_2(f-f_{\mathrm{RD}})}{\gamma_3 f_{\mathrm{damp}}}}\:,
\end{equation}
where $\{\gamma_1,\gamma_2,\gamma_3\}$ are artificial coefficients given by Eq.~(\ref{artificial_coefficients}). $\{\gamma_2,\gamma_3\}$ characterize the ringdown phase because the Lorentian part, $\gamma_3 f_\mathrm{damp}/[(f-f_\mathrm{RD})^2+{\gamma_3}^2 {f_{\mathrm{damp}}^2}]$ and the damping part $\mathrm{exp}\left[-\gamma_2(f-f_{\mathrm{RD}})/(\gamma_3 f_{\mathrm{damp}})\right]$, is motivated by an analytic behavior derived by BH perturbation theory. Thus, we can find that only $\gamma_1$, which controls the overall amplitude around the peak, characterizes the nonlinear region.

In the intermediate regime, $f_{\mathrm{a1}} \leq f \leq f_\mathrm{a2}$, a polynomial assumption 
\begin{equation}
    A_{\mathrm{int}}(f) = A_0(f)\left(\delta_0 + \delta_1 f + \delta_2 f^2 + \delta_3 f^3 + \delta_4 f^4 \right)\:,
\end{equation}
is adopted. The coefficients $\{\delta_i\}$ are 5 artificial parameters in the intermediate regime, which are fixed by the $C^1$ continuity conditions at $f=f_{\mathrm{a1}}, f_{\mathrm{a2}}$ (4 equations), and one condition at the middle frequency, $f_\mathrm{int} = (f_\mathrm{a1}+f_\mathrm{a2})/2$:
\begin{gather}
    A_\mathrm{ins}(f_\mathrm{a1}) = A_\mathrm{int}(f_\mathrm{a1})\;, \\
    A_\mathrm{int}(f_\mathrm{a2}) = A_\mathrm{MR}(f_\mathrm{a2})\;, \\ 
    A'_\mathrm{ins}(f_\mathrm{a1}) = A'_\mathrm{int}(f_\mathrm{a1})\;, \\
    A'_\mathrm{int}(f_\mathrm{a2}) = A'_\mathrm{MR}(f_\mathrm{a2})\;, \\
    A_\mathrm{int}(f_\mathrm{int}) = v_2 A_0(f_\mathrm{int})\;,
\end{gather}
where $v_2$ is an artificial coefficient determined by Eq.~\eqref{artificial_coefficients}.
We write these continuity conditions for $\{\delta_i\}$ in a matrix form,
\begin{equation}
\label{amp_connection}
\begin{pmatrix}
1 & f_\mathrm{a1} & {f_\mathrm{a1}}^2 & {f_\mathrm{a1}}^3 & {f_\mathrm{a1}}^4 \\
1 & f_\mathrm{a2} & {f_\mathrm{a2}}^2 & {f_\mathrm{a2}}^3 & {f_\mathrm{a2}}^4 \\
{} & {} & {\boldsymbol{A}_1}^\mathrm{T} & {} & {}\\
{} & {} & 
{\boldsymbol{A}_2}^\mathrm{T} & {} & {}\\
1 & f_\mathrm{int} & {f_\mathrm{int}}^2 & {f_\mathrm{int}}^3 & {f_\mathrm{int}}^4
\end{pmatrix}
\begin{pmatrix}
\delta_0 \\ \delta_1 \\ \delta_2 \\ \delta_3 \\ \delta_4 
\end{pmatrix}
=
\begin{pmatrix}
A_\mathrm{ins}(f_\mathrm{a1}) \\
A_\mathrm{MR}(f_\mathrm{a2})  \\
A'_\mathrm{ins}(f_\mathrm{a1}) \\
A'_\mathrm{MR}(f_\mathrm{a2}) \\
v_2 A_0(f_\mathrm{int})
\end{pmatrix}\;,
\end{equation}
where ${\boldsymbol{A}_{1,2}}^\mathrm{T}$ are the transposes of $\boldsymbol{A}_{1,2}$ which are defined as 
\begin{equation}
    \boldsymbol{A}_j := 
\begin{pmatrix}
\label{A}
    A'_0(f_{\mathrm{a}j}) \\ f_{\mathrm{a}j} A'_0(f_{\mathrm{a}j}) + A_0(f_{\mathrm{a}j}) \\ {f_{\mathrm{a}j}}^2 A'_0(f_{\mathrm{a}j}) +  2{f_{\mathrm{a}j}} A_0(f_{\mathrm{a}j})  \\ {f_{\mathrm{a}j}}^3 A'_0(f_{\mathrm{a}j}) +  3 {f_{\mathrm{a}j}}^2 A_0(f_{\mathrm{a}j})  \\
    {f_{\mathrm{a}j}}^4 A'_0(f_{\mathrm{a}j}) +  4 {f_{\mathrm{a}j}}^3 A_0(f_{\mathrm{a}j})
\end{pmatrix}\;,
\end{equation}
for $j=1,2$. The important point is that $\{\delta_i\}$ are determined by $v_2$ and the continuity conditions.

\subsubsection{\textbf{Nonlinear regime parameters}}
\label{non_linear_parameters}
In summary, we confirm the parameters of the nonlinear regime among $\{ \Lambda^i \}$, except for the parameters fixed by the $C^1$ continuity conditions.
\begin{itemize}
    \item 4 phase parameters, $\{ \beta_{2,3}, \alpha_{2,3} \}$
    \item 2 amplitude parameters, $\{\gamma_1, v_2\}$
\end{itemize}
We will refer to these 6 parameters as $\{ \lambda_i \}$ in the following and consider deviation from the values in GR in the next section.

\section{Modified waveform}
\label{sec:Modified_waveform}
In this section, we present our modified waveform to capture beyond-GR effects in the merger phase. Our basic idea for the waveform construction is that perturbative modifications to a GR waveform can capture deviations from GR in the nonlinear region.  
The advantage of our method is that we quantify the deviations that are difficult to calculate by
analytical and numerical methods in extended models. 
As a first implementation target, we choose the IMRPhenomD waveform.

In Sec.~\ref{sec:reparametrization} to \ref{sec:amplitude_modification}, we first present basic methods to introduce beyond-GR parameters to phase and amplitude, respectively. To introduce the parameter for phase, we consider the principal component analysis (PCA). PCA allows us to specify linearly independent components among the parameters in the nonlinear regime. We adopt the leading principal component as a parameter because other components turn out to be less significant as discussed in Sec.~\ref{sec:specification_PCA}. For amplitude, we introduce one beyond-GR parameter by modifying one of the coefficients in the IMRPhenomD waveform. This parameter describes an overall amplification in the nonlinear regime.
However, these modifications alone do not guarantee the physical consistency in the IMR process. 
Therefore, in Sec.~\ref{sec:physical_consistency} we formulate a prescription ensuring the physical consistency of the final mass and spin of a remnant BH by considering the additional losses of energy and angular momentum carried by GW radiation due to the modifications of the waveform.
Since not only the ringdown and damping frequencies but also the collocation frequencies and the continuity conditions are modified, the resultant signal is different from a signal that does not take the radiation backreactions into account. The complete form of our physically consistent waveform is shown in Sec.~\ref{sec:inclusion_backreaction}.
Finally, in Sec.~\ref{sec:physical interpretation}, the physical picture and assumptions of our work are discussed. Furthermore, in Sec.~\ref{sec:comparison}, comparisons with E.~Maggio \textit{et al.} \cite{testing_gr_Maggio-san}, which presents a parameterized waveform for a purpose similar to ours are discussed.

\begin{figure*}[t]
\includegraphics[scale=0.42]{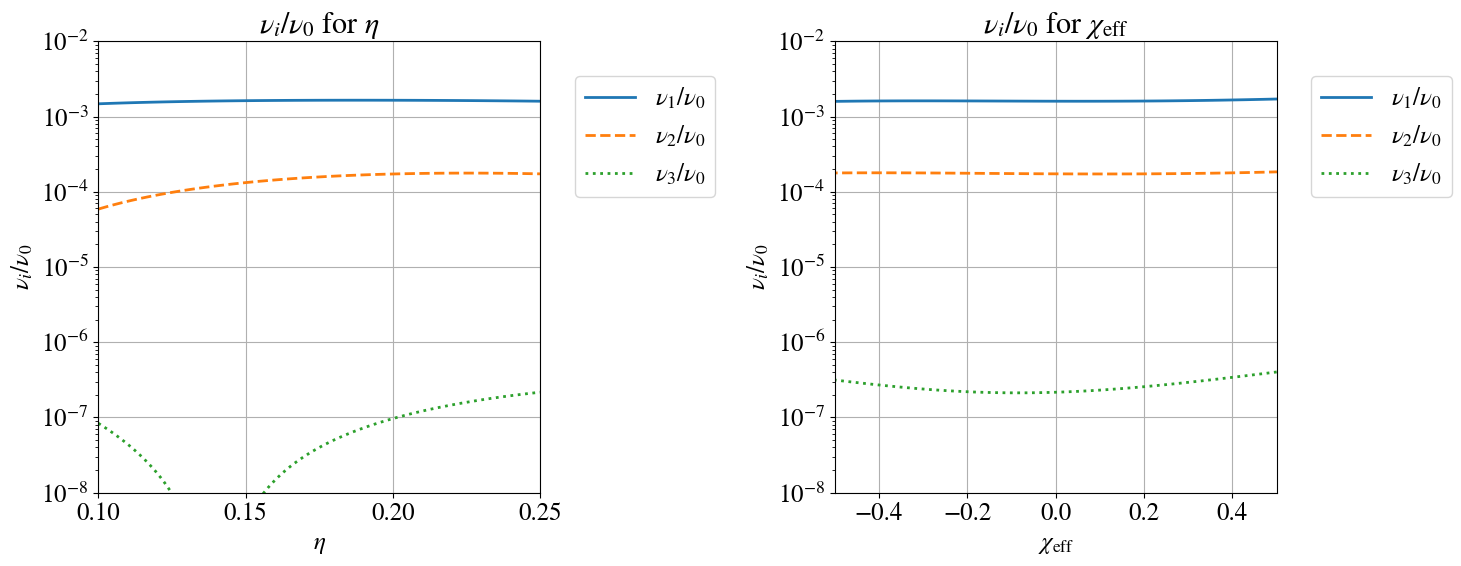}
\caption{\label{fig:eigenvalue_ratios}
Ratios of eigenvalues of $\boldsymbol{F}_{{\hat{\beta}}\hat{\alpha}}$ with $0.1 \leq \eta \leq 0.25$ fixing $\chi_\mathrm{eff}=0$. $\nu_i$ are the eigenvalues of $\boldsymbol{F}_{{\hat{\beta}}\hat{\alpha}}$, ordered from largest to smallest, $i=0,1,2,3$. The left and right figures show the fraction, $\nu_i/\nu_0$, for $\eta$ and $\chi_\mathrm{eff}$, respectively. These results show that $\nu_1/\nu_0=\mathcal{O}(10^{-3})$.}
\end{figure*}

\subsection{Re-parameterization}
\label{sec:reparametrization}
In this work, we define the fractional deviations, $\{\hat{\lambda}_i\}$, such that
\begin{equation}
    \lambda_i = \lambda_{i\_\mathrm{GR}} \left( 1 + \hat{\lambda}_i \right)\;,
\end{equation}  
where $\lambda_{i\_\mathrm{GR}}$ is the value of $\lambda_{i}$ in GR~\cite{testing_GR_GW150914, testing_GR_GWTC1, testing_GR_GWTC2}.

\subsection{Phase modification}
\label{sec:phase_modification}
As pointed out in Sec.~\ref{non_linear_parameters}, $\phi_\mathrm{GR}(f)$ has 4 parameters in the nonlinear regime, $\{\hat{\beta}_{2,3}, \hat{\alpha}_{2,3} \}$. Since these parameters are originally introduced as fitting parameters to NR waveforms, it is difficult to interpret their physical meanings explicitly. Thus, in this work, we focus on  principal components of $\{\hat{\beta}_{2,3}, \hat{\alpha}_{2,3} \}$, and expect that they play a physically important role.  We then take the leading principal component as a beyond-GR parameter for phase. 

\subsubsection{\textbf{Specification of the dominant component}}
\label{sec:specification_PCA}
To specify the leading principal component, we consider the Fisher matrix for the parameters $\{\hat{\beta}_{2,3}, \hat{\alpha}_{2,3} \}$, which is defined as
\begin{equation}
\label{Fisher_matrix}
    \boldsymbol{F}_{\hat{\beta}\hat{\alpha}} = 
    \begin{pmatrix}
    F_{\hat{\beta}_2\hat{\beta}_2} & F_{\hat{\beta}_2\hat{\beta}_3} & F_{\hat{\beta}_2\hat{\alpha}_2} & F_{\hat{\beta}_2\hat{\alpha}_3} \\
     & F_{\hat{\beta}_3\hat{\beta}_3} & F_{\hat{\beta}_3\hat{\alpha}_2} & F_{\hat{\beta}_3\hat{\alpha}_3} \\
     &  &
    F_{\hat{\alpha}_2\hat{\alpha}_2} & F_{\hat{\alpha}_2\hat{\alpha}_3} \\
    \mathrm{sym.} &  &
     & F_{\hat{\alpha}_3\hat{\alpha}_3}
    \end{pmatrix}\;,
\end{equation}
with
\begin{equation}
    \left(\boldsymbol{F}_{\hat{\beta}\hat{\alpha}}\right)_{ij} := 
    \left(\partial_{\hat{\lambda}_i} \tilde{h}(f), \partial_{\hat{\lambda}_j} \tilde{h}(f)\right)\:\mathrm{with}\:\:S_n(f)=1\;,
\end{equation}
where $\{{\hat{\lambda}}_i\} := \{ \hat{\beta}_{2, 3}, \hat{\alpha}_{2,3} \}$, and the inner product is defined in Eq.~\eqref{inner_product}. The evaluations are done at the GR values, and $S_n(f)$ is set to unity since we focus only on the structure of the waveform. Here, the lower cutoff frequency $f_\mathrm{min}$ is set to $0.0035/M$, which is the minimum frequency considered in the IMRPhenomD model.

By diagonalizing $\boldsymbol{F}_{\hat{\beta}\hat{\alpha}}$, 
 we can find the pairs of eigenvalues and eigenvectors. In principal component analysis, the eigenvalues indicate the importance of the eigenvectors. Figure~\ref{fig:eigenvalue_ratios} shows the fractions of the eigenvalues, $\nu_i/\nu_0$, where $\nu_i$ is the eigenvalue of $\boldsymbol{F}_{\hat{\beta}\hat{\alpha}}$ ordered from largest to smallest, $i=0,1,2,3$. These results show $\nu_1/\nu_0=\mathcal{O}(10^{-3})$ in the ranges of $0.1 \leq \eta \leq 0.25$ and $-0.95 \leq \chi_\mathrm{eff} \leq 0.95$. We thus adopt only $\nu_0$ as a beyond-GR parameter.

\subsubsection{\textbf{Beyond-GR parameter, $\hat{P}_\mathrm{PCA}$}}
\label{P_1,2}
Using a normalized eigenvector associated with the largest eigenvalue, $\Lambda_0$, denoted as $\Vec{e}_\mathrm{PCA}$, we define a beyond-GR parameter, $\Vec{P}_\mathrm{PCA}$, as
\begin{equation}
\label{P_PCA}
        \Vec{P}_\mathrm{PCA} = \hat{P}_\mathrm{PCA} \Vec{e}_\mathrm{PCA}\;, 
\end{equation}
where $\hat{P}_\mathrm{PCA}$ and $\Vec{e}_\mathrm{PCA}$ represent a magnitude and a direction of $\Vec{P}_\mathrm{PCA}$, respectively. We restrict $\hat{P}_\mathrm{PCA}$ to be perturbative. The fitting formula of $\Vec{e}_\mathrm{PCA}$ is given in Appendix~\ref{sec:fitting formula}.

Inversely, $\Vec{P}_\mathrm{PCA}$ corresponds to deviations from $\{\beta_{2,3}, \alpha_{2,3}\}$ in GR, that is,
\begin{gather}
    \Delta \beta_j(\hat{P}_\mathrm{PCA}) := \beta_{j\_\mathrm{GR}}\left( 1+\Vec{P}_\mathrm{PCA} \cdot{\Vec{e}_{\hat{\beta}_j}} \right)\;, \\
    \Delta \alpha_j(\hat{P}_\mathrm{PCA}) := \alpha_{j\_\mathrm{GR}}\left(1+\Vec{P}_\mathrm{PCA} \cdot{\Vec{e}_{\hat{\alpha}_j}}\right)\;,
\end{gather}
for $j=2,3$, where $\Vec{e}_{\hat{\beta}_{j}, \hat{\alpha}_j}$ are the basis vectors of $\hat{\beta}_j, \hat{\alpha}_j$ in the parameter space. 
Therefore, the deviated parameters, $\{ \beta_{j}, \alpha_{j} \}$, are
\begin{gather}
    \label{alpha_m}
    \alpha_{j} = \alpha_{j\_\mathrm{GR}} + \Delta \alpha_j(\hat{P}_\mathrm{PCA}) \;,\\
    \label{beta_m}
    \beta_{j} = \beta_{j\_\mathrm{GR}} + \Delta \beta_j(\hat{P}_\mathrm{PCA})\;,
\end{gather}
for $j=2,3$. We express the beyond-GR parameter in phase as $\hat{P}_\mathrm{PCA}$, in the following.

\subsubsection{\textbf{Modified phase}}
\label{sec:Modified phase}
\begin{figure}[t]
\includegraphics[scale=0.45]{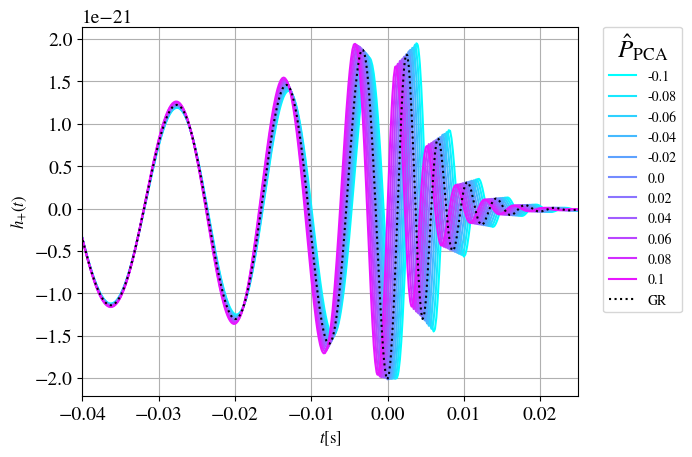}
\caption{\label{fig:P_PCA_timedomain}
Modified waveforms in the time domain associated with $\hat{P}_\mathrm{PCA}\:(|\hat{P}_\mathrm{PCA}|\leq0.1)$.}
\end{figure}

To construct the modified phase, solve a useful form of the continuity conditions, Eq.~\eqref{phase_connection}, derived by setting $\beta_j=\beta_{j\_\mathrm{GR}}+\Delta\beta_j$ and $\alpha_j=\alpha_{j\_\mathrm{GR}}+\Delta\alpha_j$ in $C_i$ (Eqs.~(\ref{C1})--(\ref{C4})), 
\begin{equation}
\label{phase_connection_modified}
\begin{pmatrix}
1 & f_\mathrm{p1} & 0 & 0\\
0 & 1 & 0 & 0\\
1 & f_\mathrm{p2} & -1 & - f_\mathrm{p2}\\
0 & 1 & 0 & -1 \\
\end{pmatrix}
\begin{pmatrix}
\Delta \beta_0 \\ \Delta \beta_1 \\ \Delta \alpha_0 \\ \Delta \alpha_1
\end{pmatrix}
= 
\begin{pmatrix}
\Delta C_1 \\ \Delta C_2 \\ \Delta C_3 \\ \Delta C_4\\
\end{pmatrix}\;,
\end{equation}
with
\begin{gather}
    \Delta C_1 =  -\log{(f_\mathrm{p1})}\Delta\beta_2 + \frac{1}{3} {f_\mathrm{p1}}^{-3}\Delta\beta_3\;, \label{dC1}\\   
    \Delta C_2 = - {f_\mathrm{p1}}^{-1}\Delta\beta_2 -  {f_\mathrm{p1}}^{-3}\Delta\beta_3\;,  \label{dC2}  \\  
    \begin{split}
    \Delta C_3 = & - {f_\mathrm{p2}}^{-1}\Delta\alpha_2 + \frac{4}{3} {f_\mathrm{p2}}^{3/4}\Delta\alpha_3\\ &-  \log{f_\mathrm{p2}}\Delta\beta_2 + \frac{1}{3}{f_\mathrm{p2}}^{-3}\Delta\beta_3\;,\label{dC3} \\ 
    \end{split}\\
    \begin{split}
    \Delta C_4 = &{f_\mathrm{p2}}^{-2}\Delta\alpha_2 + {f_\mathrm{p2}}^{-1/4}\Delta\alpha_3\\ &- {f_\mathrm{p2}}^{-1}\Delta\beta_2 -  {f_\mathrm{p2}}^{-4}\Delta\beta_3\;, \label{dC4}
    \end{split}
\end{gather}
where $\{\beta_{k}, \alpha_{k} \}$ for $k=0,1$ are deviated from $\{\beta_{k\_\mathrm{GR}}, \alpha_{k\_\mathrm{GR}} \}$. 
Then, we can construct the modified phase
\begin{equation}
\label{modified_phase}
\phi_{\mathrm{m}}(f;\hat{P}_\mathrm{PCA})=
\begin{cases}
\phi_{\mathrm{ins}}(f), & f\geq f_{\mathrm{p1}} \\
\phi_{\mathrm{int}\_\mathrm{m}}(f;\hat{P}_\mathrm{PCA}), & f_{\mathrm{p1}} \leq f \leq f_{\mathrm{p2}} \\
\phi_{\mathrm{MR}\_\mathrm{m}}(f;\hat{P}_\mathrm{PCA}), & f \geq f_{\mathrm{p2}}
\end{cases}\;,
\end{equation}
where $\phi_{\mathrm{int}\_\mathrm{m}}(f;\hat{P}_\mathrm{PCA}),\phi_{\mathrm{MR}\_\mathrm{m}}(f;\hat{P}_\mathrm{PCA})$ are the modified phases for the intermediate and merger-ringdown parts due to $\hat{P}_\mathrm{PCA}$, respectively. Figure~\ref{fig:P_PCA_timedomain} shows that $\hat{P}_\mathrm{PCA}$  shifts the time and phase in the nonlinear regime compared to those of the GR waveform. Note that the amplification here is caused by the coalescence time and phase shifts. 
not by a deviation in amplitude. The quantitative forms of these shifts can be derived by considering the continuity conditions.
The derivation of these shifts is shown in Appendix~\ref{sec:time and phase shifts}. 
In Figure~\ref{fig:P_PCA_timedomain} and subsequent figures in this section, GW150914-like parameters shown in Table~\ref{tb:binary} are used for illustrations. The redshift $z$, which scales the total mass dependence in the GW signal, is neglected here. 

\begin{table}[h]
\caption{\label{tb:binary}%
GW150914-like parameters used for 
illustrations
}
\begin{ruledtabular}
\begin{tabular}{ccccccc}
Binary & $M\:(M_\odot)$ & $\eta$ & $\chi_1$ & $\chi_2$ & $r\:$(Mpc) & $\iota$ \\
\colrule
{} & 68 & 0.25 & 0 & 0 & 400 & 0 \\
\end{tabular}
\end{ruledtabular}
\end{table}

\subsection{Amplitude modification}
\label{sec:amplitude_modification}
While the amplitude $A_\mathrm{GR}(f)$ has two parameters, $\gamma_1$ and $ v_2$, in the nonlinear regime, we employ one beyond-GR parameter in amplitude, denoted by $\hat{\gamma}_1$, instead of specifying principal components as well as done for phase. This is because we assume that $ v_2$ linearly depends on $\gamma_1$, as we show explicitly in Sec.~\ref{sec:hat_gamma_1}. The assumption is motivated by the fact that nonlinearity of gravity would increase toward the merger of a compact binary.

\subsubsection{\textbf{Beyond-GR parameter,  $\hat{\gamma}_1$}}\label{sec:hat_gamma_1}
Naively, $\hat{\gamma_1}$ is a parameter that describes an overall modification for amplitude in the merger-ringdown part as discussed in Sec.~\ref{sec:PhenomD_amplitude}. Then, we define the modified amplitude in the merger-ringdown part as
\begin{equation}
\begin{split}
    A_{\mathrm{MR}\_\mathrm{m}}(f;\hat{\gamma}_1) &:= (1+\hat{\gamma}_1)A_\mathrm{MR}(f)\\
    &= A_\mathrm{MR}(f) + \Delta A_\mathrm{MR}(f;\hat{\gamma}_1)\;,
\end{split}
\end{equation}
where
\begin{equation}
\label{DeltaA_MR}
    \Delta A_\mathrm{MR}(f;\hat{\gamma}_1) := \hat{\gamma}_1 A_\mathrm{MR}(f)\;.
\end{equation}
In addition, we assume the linear frequency dependence of amplitude modification in the intermediate part by interpolating amplitudes at $f_{a1}$ and $f_{a2}$. This is achieved by setting $v_2$ at the middle frequency, $f_{\mathrm{int}}$, as
\begin{equation}
    v_{2} := v_{2\_\mathrm{GR}} + \Delta v_2(\hat{\gamma}_1)\;,
\end{equation}
where
\begin{equation}
\label{Deltav_2}
    \Delta v_2(\hat{\gamma}_1) := \frac{1}{2}\Delta A_{\mathrm{MR}}(f_{a2};\hat{\gamma}_1)\;.
\end{equation}

\subsubsection{\textbf{Modified amplitude}}
Fixing an initial configuration and $\hat{\gamma}_1$, we can determine the intermediate part
\begin{equation}
    A_{\mathrm{int}}(f;\hat{\gamma}_1) =  A_0(f)\sum\limits_{i=0}^{4}\delta_{i}f^{i}\;,
\end{equation}
by solving the modified form of Eq.~(\ref{amp_connection})
\begin{equation}
\label{amp_connection_modified}
\begin{split}
&
\begin{pmatrix}
1 & f_\mathrm{a1} & {f_\mathrm{a1}}^2 & 
{f_\mathrm{a1}}^3 & {f_\mathrm{a1}}^4 \\
1 & f_\mathrm{a2} & {f_\mathrm{a2}}^2 & {f_\mathrm{a2}}^3 & {f_\mathrm{a2}}^4 \\
{} & {} & {\boldsymbol{A}_1}^\mathrm{T} & {} & {}\\
{} & {} & {\boldsymbol{A}_2}^\mathrm{T} & {} & {}\\
1 & f_\mathrm{int} & {f_\mathrm{int}}^2 & {f_\mathrm{int}}^3 & {f_\mathrm{int}}^4
\end{pmatrix}
\begin{pmatrix}
\delta_{0} \\ \delta_{1} \\ \delta_{2} \\ \delta_{3} \\ \delta_{4} 
\end{pmatrix}\\
&\:\:\:\:\:\:=
\begin{pmatrix}
A_\mathrm{ins}(f_\mathrm{a1}) \\
A_{\mathrm{MR}\_\mathrm{m}}(f_\mathrm{a2};\hat{\gamma}_1)  \\
A'_\mathrm{ins}(f_\mathrm{a1}) \\
A'_{\mathrm{MR}\_\mathrm{m}}(f_\mathrm{a2};\hat{\gamma}_1) \\
v_{2}(\hat{\gamma}_1)\;,
\end{pmatrix}.
\end{split}
\end{equation}
where $\{\delta_{i}\}$ implicitly depend on $\{\hat{\gamma}_1\}$.

\begin{figure}[t]
\includegraphics[scale=0.45]{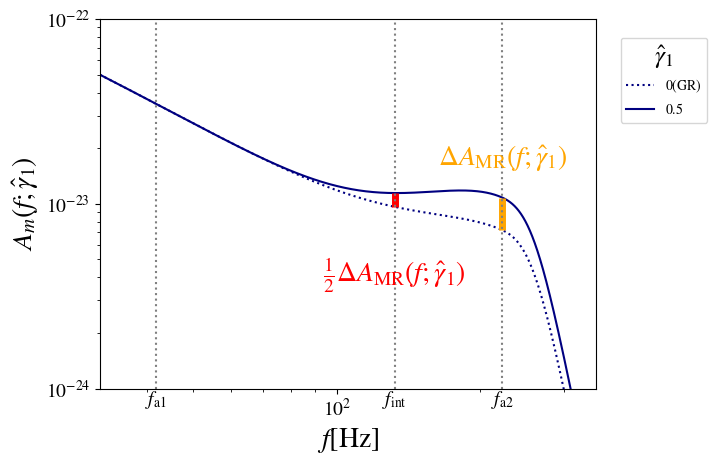}
\caption{\label{fig:gamma1 freq domain}
Modified amplitude with $\hat{\gamma}_1 = 0.5$ (navy solid line) and GR amplitude (navy dotted line). $\hat{\gamma}_1$ describes the amplification at $f=f_\mathrm{a2}$ (orange) and the linear interpolation at $f=f_\mathrm{int}$ (red). The former and latter deviations correspond to Eqs.~\eqref{DeltaA_MR} and \eqref{Deltav_2}, respectively.}
\end{figure}

From these procedures, we can construct the modified amplitude
\begin{equation}
A_{\mathrm{m}}(f;\hat{\gamma}_1)=
\begin{cases}
A_{\mathrm{ins}}(f),  & f\leq f_{\mathrm{a1}} \\
A_{\mathrm{int}\_\mathrm{m}}(f;\hat{\gamma}_1),  & f_{\mathrm{a1}} \leq f \leq f_{\mathrm{a2}} \\
A_{\mathrm{MR}\_\mathrm{m}}(f;\hat{\gamma}_1),  & f \geq f_{\mathrm{a2}}\;
\end{cases}\;.
\label{modified_amplitude}
\end{equation}
Figure~\ref{fig:gamma1 freq domain} shows how the modified amplitude in the frequency domain.

\subsection{Physical consistency}
\label{sec:physical_consistency}
From Eqs.~(\ref{modified_phase}) and (\ref{modified_amplitude}), we can construct a modified waveform
\begin{equation}
\label{modified_waveform}
    \tilde{h}_\mathrm{m}(f;\hat{\gamma}_1,\hat{P}_\mathrm{PCA}):=A_\mathrm{m}(f;\hat{\gamma}_1)\mathrm{e}^{-i\phi_\mathrm{m}(f;\hat{P}_\mathrm{PCA})}\;.
\end{equation}
The waveform is physically inconsistent in the sense that it does not include the radiation backreaction caused by the newly introduced parameters. In other words, $\tilde{h}_\mathrm{m}(f)$ describes a physically unnatural situation in which the properties of a remnant BH are same as those in GR, although there are additional losses of energy and angular momentum due to modifications in the nonlinear regime. 
Therefore, to create a physically consistent waveform, we need to include the effects. This goal is achieved by the following two steps.
The first step is calculating the additional energy and angular momentum losses due to the existence of beyond-GR parameters. 

The second step is reflecting these deviations on the remnant BH spin, $a_\mathrm{f}$. While the additional radiation causes deviations in the mass and spin of a remnant BH, we focus only on deviation in $a_\mathrm{f}$ because Eqs.~\eqref{formula_f_RD}- \eqref{formula_f_damp}, that are used for the modeling of the ringdown GW signal, depend on $a_\mathrm{f}$. 

Deviation in $a_\mathrm{f}$ changes the ringdown GW signal because the ringdown and damping frequencies, $f_\mathrm{RD}$ and $f_\mathrm{damp}$, are modified.
In addtition, other quantities that depend on $a_\mathrm{f}$, such as the collocation frequencies, $f_\mathrm{p2}$ and $f_\mathrm{a2}$, are modified. For usefulness to implement, we revive the dependence of $c$ and $G$. 

\subsubsection{\textbf{Formulas for energy and angular momentum losses}}
First of all, we derive formulas for energy and angular momentum losses for a waveform in the frequency domain. We 
start with a waveform in the time domain, related to outgoing GW radiation, 
\begin{equation}
\label{H}
    H(t) := h_+(t) - ih_\times (t) = \sum_{l=2}^{\infty} \sum^{l}_{m=-l}  {}^{-2}Y_{lm}(\theta, \phi) h_{lm}(t)\;,
\end{equation}
where ${}^{-2}Y_{lm}(\theta,\phi)$ is  the spin-weighted spherical harmonics for $s=-2$.
In the time domain, the emission rates of energy and the $z$ component of angular momentum carried by GWs (positive values mean losses from a system) are written as 
\begin{gather}
    \frac{\diff E}{ \diff t}  =  \frac{c^3 r^2}{16\pi G}\int \diff \Omega \: |\dot{H}|^2\;, \\
    \frac{\diff J^z}{\diff t} = -\frac{c^3 r^2}{16\pi G}\: 
    \mathrm{Re} \left[ \int \diff \Omega \: \frac{\partial {H}}{\partial \phi} \dot{{H}^\ast} \right]\;,
\end{gather}
respectively, where $\diff \Omega$ is the standard solid angle element \cite{ruiz_2008}. Thus, the total amounts of radiated energy and angular momentum during the coalescence are 
\begin{gather}
\label{dE}
    \Delta E = \frac{c^3 r^2}{16\pi G} \int^{\infty}_{-\infty} \diff t \int \diff \Omega \: |\dot{H}|^2 \;, \\
\label{dJz}
    \Delta J^z = -\frac{c^3 r^2}{16\pi G} \: 
    \mathrm{Re} \left[ \int^{\infty}_{-\infty} \diff t  \int \diff \Omega \: \frac{\partial {H}}{\partial \phi} \dot{{H}^\ast} \right]\;.
\end{gather}
Note that Eqs.~(\ref{dE}) and (\ref{dJz}) are asymptotic expressions ignoring the higher order corrections above $1/r^3$. 

We consider only the dominant GW modes, $(l,m)=(2,\pm 2)$, here. 
Replacing $H(t)$ with the inverse Fourier transform,
\begin{gather}
    \tilde{H}(f) = \tilde{h}_+(f) -i \tilde{h}_\times(f) = \sum_{l=2}^{\infty} \sum^{l}_{m=-l} {}^{-2}Y_{lm}(\theta, \phi) \tilde{h}_{lm}(f)\;,
\end{gather}
and integrating with respect to $t$, the total amount of radiated energy and angular momentum are 
\begin{gather}
    \label{DeltaE}
    \left( {\Delta E}\right)_\mathrm{GR}  = \frac{2 \pi c^3 r^2}{G} \int^{\infty}_{0} \diff f \:f^2 {A^2_{\mathrm{GR}}(f)}\;, \\
    \label{DeltaJz}
    \left({\Delta J^z}\right)_\mathrm{GR} = \frac{2 c^3 r^2}{G} \int^\infty_{0} \diff f \: fA^2_\mathrm{GR}(f)\;.
\end{gather}
Detailed derivations of Eqs.~\eqref{DeltaE} and \eqref{DeltaJz} are given in Appendix~\ref{sec:derivation of formulas}.
These show that radiated energy and angular momentum are completely determined by the spectrum of GW amplitude in this formalism.

\subsubsection{\textbf{Backreaction to a remnant BH}}
\label{sec:dev_remnant}
As long as perturbative modifications on a GR waveform are considered, it is expected that the radiated energy and angular momentum of $\tilde{h}_\mathrm{m}(f)$, denoted by $\left(\Delta E\right)_\mathrm{m}$ and $\left(\Delta J^z\right)_\mathrm{m}$, are written as
\begin{gather}
    \left( {\Delta E}\right)_\mathrm{m}(\hat{\gamma}_1)  = \frac{2 \pi c^3 r^2}{G} \int^{\infty}_{0} \diff f \:f^2 {A^2_{\mathrm{m}}(f;\hat{\gamma}_1)}, \\
    \left({\Delta J^z}\right)_\mathrm{m}(\hat{\gamma}_1) = \frac{2 c^3 r^2}{G} \int^\infty_{0} \diff f \: fA^2_\mathrm{m}(f;\hat{\gamma}_1)\;.
\end{gather}
Then, deviations in the radiated energy and angular momentum are
\begin{gather}
    \begin{split}
    \label{DE_dev}
        \delta \left( \Delta E \right)(\hat{\gamma}_1) &:= \left( {\Delta E}\right)_\mathrm{m}(\hat{\gamma}_1) - \left( {\Delta E}\right)_\mathrm{GR} \\
        &= \frac{2 \pi c^3 r^2}{G} \int^{\infty}_{f_\mathrm{a1}} \diff f \:f^2 \left(A^2_{\mathrm{m}}(f;\hat{\gamma}_1) - {A^2_{\mathrm{GR}}(f)}\right)\;,
    \end{split} \\
    \begin{split}
    \label{DJ_dev}
        \delta \left( \Delta J^z \right)(\hat{\gamma}_1) &:= \left( {\Delta E}\right)_\mathrm{m}(\hat{\gamma}_1) - \left( {\Delta E}\right)_\mathrm{GR} \\
        &= \frac{2 c^3 r^2}{G} \int^{\infty}_{f_\mathrm{a1}} \diff f \:f \left(A^2_{\mathrm{m}}(f;\hat{\gamma}_1) - {A^2_{\mathrm{GR}}(f)}\right)\;.
    \end{split}
\end{gather}
The lower frequency in the integral is set to $f_\mathrm{a1}$ because the deviation occurs in $f\geq f_\mathrm{a1}$.
These deviations change the final mass and spin to
\begin{gather}
    \bar{M}_\mathrm{f}(\hat{\gamma}_1) = M_\mathrm{f} - \frac{\delta(\Delta E)(\hat{\gamma}_1)}{c^2}\;, \\
    \bar{S}_\mathrm{f}(\hat{\gamma}_1) = S_\mathrm{f}-\delta(\Delta J^z)(\hat{\gamma}_1)\;.
\end{gather}
where $M_\mathrm{f}$ and $S_\mathrm{f}$ are the final mass and spin predicted by GR. Similar to $\bar{M}_\mathrm{f}$ and $\bar{S}_\mathrm{f}$, we denote the quantities that include the radiation backreactions with bar in the following.

\begin{figure*}[t]
\includegraphics[scale=0.45]{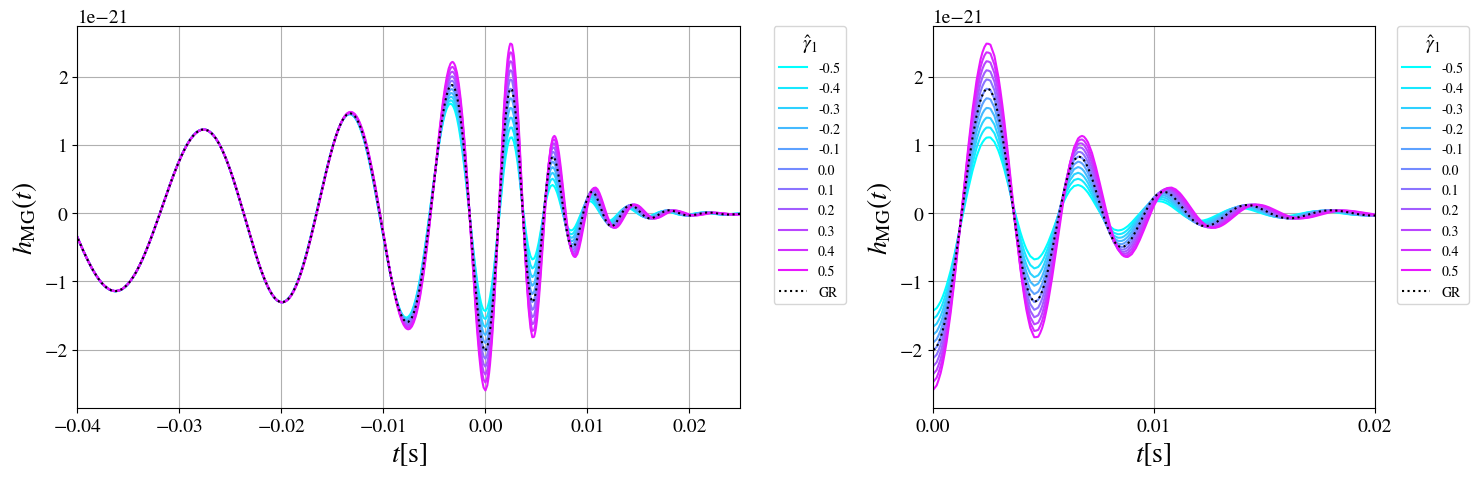}
\caption{\label{fig:gamma1 time domain}
Physically consistent modified waveforms in the time domain associated with $\hat{\gamma}_1$ ($|\hat{\gamma}_1|\leq 0.5$), that describes an amplification in the nonlinear regime. The left figure shows the IMR waveforms and the right shows the ringdown parts. Due to the backreaction inclusion associated with $\hat{\gamma}_1$, the ringdown and damping frequencies are changed. These changes are studied in Fig.~\ref{fig:deviations on freq}.}
\end{figure*}

\begin{figure*}[t]
\includegraphics[scale=0.5]{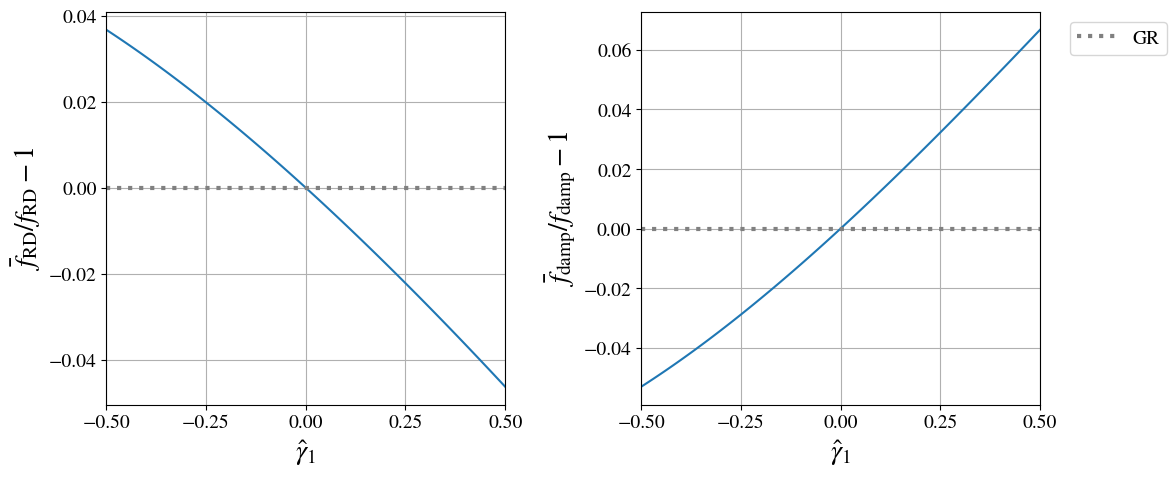}
\caption{\label{fig:deviations on freq}
Fractional deviations of the ringdown frequency (the left) and the damping frequency (the right) associated with $\hat{\gamma}_1$. $\bar{f}_\mathrm{RD}-f_\mathrm{RD}$ and $\bar{f}_\mathrm{damp}-f_\mathrm{damp}$ are monotonically decreasing and increasing functions of $\hat{\gamma}_1$, respectively.} 
\end{figure*}

\subsection{\textbf{Complete form of modified waveform}}
\label{sec:inclusion_backreaction}
The important quantity for ringdown modeling is the dimensionless Kerr parameter, $a_\mathrm{f}$, which is defined in Eq.~\eqref{a_f}.
$a_\mathrm{f}$ determines not only the ringdown and damping frequencies, $f_\mathrm{RD}$ and $f_\mathrm{damp}$, via Eqs.~(\ref{formula_f_RD}-\ref{formula_f_damp}), but also the collocation frequencies, $f_\mathrm{a2}$ and $f_\mathrm{p2}$, and the continuity conditions. Regarding these facts, our strategy to create a physically consistent waveform is as follows: 
\begin{enumerate}
    \item Fixing $\hat{\gamma}_1$, construct $\tilde{h}_\mathrm{m}(f;\hat{\gamma}_1)$, then calculate $\bar{a}_\mathrm{f}$.
    \item Replacing $a_\mathrm{f}$ with $\bar{a}_\mathrm{f}$, modify the ringdown part of the signal. 
    \item Construct an IMR waveform considering $\hat{P}_\mathrm{PCA}$ and the modified continuity conditions associated with the change of $a_\mathrm{f}$ to $\bar{a}_\mathrm{f}$.
\end{enumerate}
The first step is done, following Sec.~\ref{sec:dev_remnant}. As the second step, since $\bar{a}_\mathrm{f}$ modifies the ringdown and damping frequencies as $\bar{f}_\mathrm{RD}$ and $\bar{f}_\mathrm{damp}$, we modify phase as
\begin{equation}
\begin{split}
\label{bar_phi_MR}
&\bar{\phi}_{\mathrm{MR}\_\mathrm{m}}(f;\hat{\gamma}_1, \hat{P}_\mathrm{PCA}) \\
    &
    \begin{split}
        :=\frac{1}{\eta} \biggl\{ &\alpha_{0} + \alpha_{1} f - \alpha_{2}(\hat{P}_\mathrm{PCA}) f^{-1} + \frac{4}{3} \alpha_{3}(\hat{P}_\mathrm{PCA}) f^{3/4} \\ & + \alpha_{4\_\mathrm{GR}}\tan^{-1}{ \biggl( \frac{f-\alpha_{5\_\mathrm{GR}} \bar{f}_\mathrm{RD}}{\bar{f}_\mathrm{damp}}} \biggr) \biggr\}\;,
    \end{split} 
\end{split}
\end{equation}
and amplitude as
\begin{equation}
\begin{split}
    &\bar{A}_\mathrm{m}(f;\hat{\gamma}_1)\\ &:= (1+\hat{\gamma}_1) \frac{\gamma_{1\_\mathrm{GR}}\gamma_{3\_\mathrm{GR}} \bar{f}_{\mathrm{damp}}A_0(f)}{(f-\bar{f}_{\mathrm{RD}})^2+ {\gamma_{3\_\mathrm{GR}}}^2 {\bar{f}_{\mathrm{damp}}}^2  } \mathrm{e}^{-\frac{\gamma_{2\_\mathrm{GR}}(f-\bar{f}_{\mathrm{RD}})}{\gamma_{3\_\mathrm{GR}} \bar{f}_{\mathrm{damp}}}}\;.
\end{split}
\end{equation}
Figure~\ref{fig:gamma1 time domain} shows the physicallly consistent waveforms in the time domain associated with $\hat{\gamma}_1$ ($|\hat{\gamma}_1|\leq 0.5$). The left figure shows the IMR waveforms and the right shows the ringdown parts. Due to the backreaction inclusion associated with $\hat{\gamma}_1$, the ringdown and damping frequencies are changed.
Figure~\ref{fig:deviations on freq} shows fractional deviations of the ringdown and damping frequencies associated with the variation of $\hat{\gamma}_1$, respectively.  
On the left-hand side of Eq.~\eqref{bar_phi_MR}, we emphasize the implicit dependence of $\hat{\gamma}_1$.
For the third step, we consider changes in the collocation frequencies and the connection conditions. The collocation frequencies between the intermediate and merger-ringdown parts,
$f_\mathrm{a2}$ and $f_\mathrm{p2}$, are modified to
\begin{gather}
    \bar{f}_\mathrm{p2} = 0.5\bar{f}_\mathrm{RD} \;,\\
    \bar{f}_\mathrm{a2} := \bar{f}_\mathrm{peak} := \left| \bar{f}_\mathrm{RD} - \frac{\gamma_3 \bar{f}_{\mathrm{damp}}\left(1-\sqrt{1-{\gamma_2}^2}\right)}{\gamma_2} \right|\;.
\end{gather}
Moreover, phase for the intermediate is changed to
\begin{equation}
\label{bar_phi_int}
\begin{split}
    &\bar{\phi}_{\mathrm{int}}(f;\hat{P}_\mathrm{PCA})\\ &
    \begin{split}
        := \frac{1}{\eta}
        \Bigl( &\beta_{0} + \beta_{1} f + 
        \beta_{2}(\hat{P}_\mathrm{PCA}) \log{f} - \frac{\beta_{3}(\hat{P}_\mathrm{PCA})}{3} f^{-3} \Bigr)\;,  
    \end{split}  
\end{split}
\end{equation}
where $\{ \beta_{0},\beta_{1}, \alpha_{0}, \alpha_{1} \}$ are determined by modified connection
conditions, which are given as
\begin{equation}
\label{phase_connection_modified2}
\begin{pmatrix}
1 & f_\mathrm{p1} & 0 & 0\\
0 & 1 & 0 & 0\\
1 & \bar{f}_\mathrm{p2} & -1 & - \bar{f}_\mathrm{p2}\\
0 & 1 & 0 & -1 \\
\end{pmatrix}
\begin{pmatrix}
\Delta \beta_0 \\ \Delta \beta_1 \\ \Delta \bar{\alpha}_0 \\ \Delta \bar{\alpha}_1
\end{pmatrix}
= 
\begin{pmatrix}
\Delta C_1 \\ \Delta C_2 \\ {\Delta \bar{C}_3} \\ {\Delta \bar{C}_4}\\
\end{pmatrix}\;,
\end{equation}
where $\Delta \bar{C}_{3,4}$ are obtained by replacing $f_\mathrm{p2}$ in Eq.~(\ref{dC3}) and (\ref{dC4}) with $\bar{f}_\mathrm{p2}$. Since we keep the inspiral part the same as GR, these changes affect only $\Delta \bar{\alpha}_{0.1}$.
Similarly, for amplitude, $\bar{A}_{\mathrm{int}\_\mathrm{m}}(f;\hat{\gamma}_1)$ becomes
\begin{equation}
   \bar{A}_{\mathrm{int}\_\mathrm{m}}(f;\hat{\gamma}_1) := A_0(f)\sum^4_{j=0}\bar{\delta}_{i}f^i\;,
\end{equation}
where $\{\bar{\delta}_{i}\}$ are determined by the modified connection conditions,
\begin{equation}
\label{amp_m_connection}
\begin{split}
&
\begin{pmatrix}
1 & f_\mathrm{a1} & {f_\mathrm{a1}}^2 & 
{f_\mathrm{a1}}^3 & {f_\mathrm{a1}}^4 \\
1 & \bar{f}_\mathrm{a2} & {\bar{f}_\mathrm{a2}}^2 & {\bar{f}_\mathrm{a2}}^3 & {\bar{f}_\mathrm{a2}}^4 \\
{} & {} & {\boldsymbol{A}_1}^\mathrm{T} & {} & {}\\
{} & {} & {\bar{\boldsymbol{A}_2}}^\mathrm{T} & {} & {}\\
1 & \bar{f}_\mathrm{int} & {\bar{f}_\mathrm{int}}^2 & {\bar{f}_\mathrm{int}}^3 & {\bar{f}_\mathrm{int}}^4
\end{pmatrix}
\begin{pmatrix}
\bar{\delta}_{0} \\ \bar{\delta}_{1} \\ \bar{\delta}_{2} \\ \bar{\delta}_{3} \\ \bar{\delta}_{4} 
\end{pmatrix} \\
&\:\:\:\:\:\:=
\begin{pmatrix}
A_\mathrm{ins}(f_\mathrm{a1}) \\
\bar{A}_{\mathrm{MR}\_\mathrm{m}}(\bar{f}_\mathrm{a2};\hat{\gamma}_1)  \\
A'_\mathrm{ins}(f_\mathrm{a1}) \\
{\bar{A}}'_{\mathrm{MR}\_\mathrm{m}}(\bar{f}_\mathrm{a2};\hat{\gamma}_1) \\
v_{2\_\mathrm{m}}(\hat{\gamma}_1) A_0(\bar{f}_\mathrm{int})
\end{pmatrix}\;,
\end{split}
\end{equation}
where $\bar{f}_\mathrm{int}:=(f_\mathrm{a1}+\bar{f}_\mathrm{a2})/2$ and $\bar{\boldsymbol{A}}_2$ is obtained by replacing $f_\mathrm{p2}$ in Eq.~(\ref{A}) with $\bar{f}_\mathrm{p2}$.
Then, we finally derive the complete form of our waveform
\begin{equation}
\label{modified_waveform_final}
\tilde{h}_\mathrm{MG}(f;\hat{\gamma}_1,\hat{P}_\mathrm{PCA}) := {A}_\mathrm{MG}(f;\hat{\gamma}_1)\mathrm{e}^{-i{\phi}_\mathrm{MG}(f;\hat{\gamma}_1, \hat{P}_\mathrm{PCA})}\;,
\end{equation}
where
\begin{equation}
\label{phi_m}
\begin{split}
&{\phi}_\mathrm{MG}(f;\hat{\gamma}_1, \hat{P}_\mathrm{PCA}) \\ &:= 
\begin{cases}
\phi_{\mathrm{ins}}(f), & f \leq f_{\mathrm{p1}}\\
\bar{\phi}_{\mathrm{int\_m}}(f;\hat{P}_{\mathrm{PCA}}), & f_{\mathrm{p1}} \leq f \leq \bar{f}_{\mathrm{p2}}\\
\bar{\phi}_{\mathrm{MR\_m}}(f;\hat{\gamma}_1, \hat{P}_{\mathrm{PCA}}), & f \geq \bar{f}_{\mathrm{p2}}
\end{cases}
\;,
\end{split}
\end{equation}
\begin{equation}
\label{A_m}
{A}_\mathrm{MG}(f;\hat{\gamma}_1) := 
\begin{cases}
{A}_{\mathrm{ins}}(f), & f \leq f_{\mathrm{a1}}\\
\bar{A}_{\mathrm{int\_m}}(f;\hat{\gamma}_1), & f_{\mathrm{a1}} \leq f \leq \bar{f}_{\mathrm{a2}}\\
\bar{A}_{\mathrm{MR\_m}}(f;\hat{\gamma}_1), & f \geq \bar{f}_{\mathrm{a2}}
\end{cases}
\;.
\end{equation}
  
\subsection{\textbf{Availability and assumption of our waveform}}
\label{sec:physical interpretation}
Since our prescription to ensure the physical consistency does not depend on specific models, it is expected to be applicable to various cases of the nonstandard models of BBH mergers, such as extended theories of gravity (e.g,~\cite{Berti2018}), or environmental effects of BHs (e.g.,~\cite{Fedrow_2017}). To investigate the compatibility of our waveform with those in specific models of extended gravity theories, in Sec.~\ref{sec:EdGB}, we show that our waveform can reproduce Einstein-dilaton Gauss-Bonnet gravity waveforms~\cite{EdGB_waveform} within the measurement errors in the O4 and O5 observations.

To keep our waveform physically consistent, we need to work within the perturbative regime from the
original IMRPhenomD waveform. First, we demand that beyond-GR modifications on the waveform are perturbative. Under this requirement, it is strongly expected that the modification to $a_\mathrm{f}$ is also perturbative.
Indeed, for $|\hat{\gamma}_1| \lesssim 0.4$,  a relative error of the dimensionless Kerr parameter $a_\mathrm{f}$ between estimated from the complete waveform, $\tilde{{h}}_\mathrm{MG}(f)$, and from the tentative waveform (without the backreaction), $\tilde{h}_\mathrm{m}(f)$, is $\leq 3 \%$, which is comparable to a systematic error for the modeling of $a_\mathrm{f}$~\cite{PhysRevD.95.064024}. On the other hand, there is no explicit indicator to assess the regime of validity for $\hat{P}_\mathrm{PCA}$ since at the waveform generation level, $\hat{P}_\mathrm{PCA}$ can take arbitrary value because the connection procedure (to impose C${}^1$ condition) works for any value of $\hat{P}_\mathrm{PCA}$. However, too large values of $\hat{P}_\mathrm{PCA}$ imply non-perturbative modifications to the BBH dynamics. Therefore, we impose that $\hat{P}_\mathrm{PCA}$ must take a value in a perturbative range because we are here interested in perturbative deviations from GR.

\subsection{Comparison with Maggio \textit{et al.}~\cite{testing_gr_Maggio-san}}
\label{sec:comparison}
Finally, we compare our waveform model with the previous work by Maggio \textit{et al.}~\cite{testing_gr_Maggio-san}, who propose a parameterized waveform model that can capture the deviations in the merger and the ringdown regimes. They introduce 5 beyond-GR parameters for each GW mode $(l,m)$, $\{ \delta A_{lm}, \delta \omega_{lm}, \delta \Delta t_{lm}, \delta f_{lm0}, \delta \tau_{lm0} \}$, that characterize the merger and ringdown parts. The parameters, $\{ \delta A_{lm}, \delta \omega_{lm}, \delta \Delta t_{lm} \}$, are ones for the merger waveform and describe fractional deviations in amplitude, angular frequency, and time-lag, respectively. The other two parameters, $\{ \delta f_{lm0}, \delta \tau_{lm0} \}$, are ones for the ringdown waveform and describe fractional deviations in the ringdown frequency and damping time for the fundamental mode, $n=0$, respectively. Both our and their waveforms parameterize the deviation from GR in the nonlinear region.

First, a main difference from \cite{testing_gr_Maggio-san} is that our waveform is modeled in the frequency domain, while their waveform is in the time domain. Their waveform model is constructed considering generic deviations in the merger and ringdown regimes based on the EOB waveform~\cite{bohe_2017, cotesta_2018}. 
Another point is that considering only the quadrupole GW modes, our parameterization corresponds to picking 2 degrees of freedom from 3 parameters characterizing the nonlinear regime in their model, $\{\delta A_{220}, \delta \omega_{22}, \delta \Delta t_{22} \}$. For phase, we adopt only one parameter, $\hat{P}_\mathrm{PCA}$, which determines time and phase shifts dependently, while \cite{testing_gr_Maggio-san} introduces those shifts independently. This is because one of our purposes in this study is to specify the dominant components among the artificial parameters introduced in the IMRPhenomD waveform. 

Furthermore, by including the radiation reactions associated with modifications in the nonlinear regime, to the remnant, we give an implicit but physical relation between $\delta A_{220}$, and $\delta f_{220}$ and $\delta \tau_{220}$ that characterize the ringdown regime in the model. In other words, our procedure to include the reaction removes potential degeneracies between deviations in amplitude and in the ringdown and damping frequencies in a physically-consistent way. We emphasize again that in this study we do not introduce free parameters in the ringdown part. 

\begin{figure*}[t]
\includegraphics[scale=0.42]{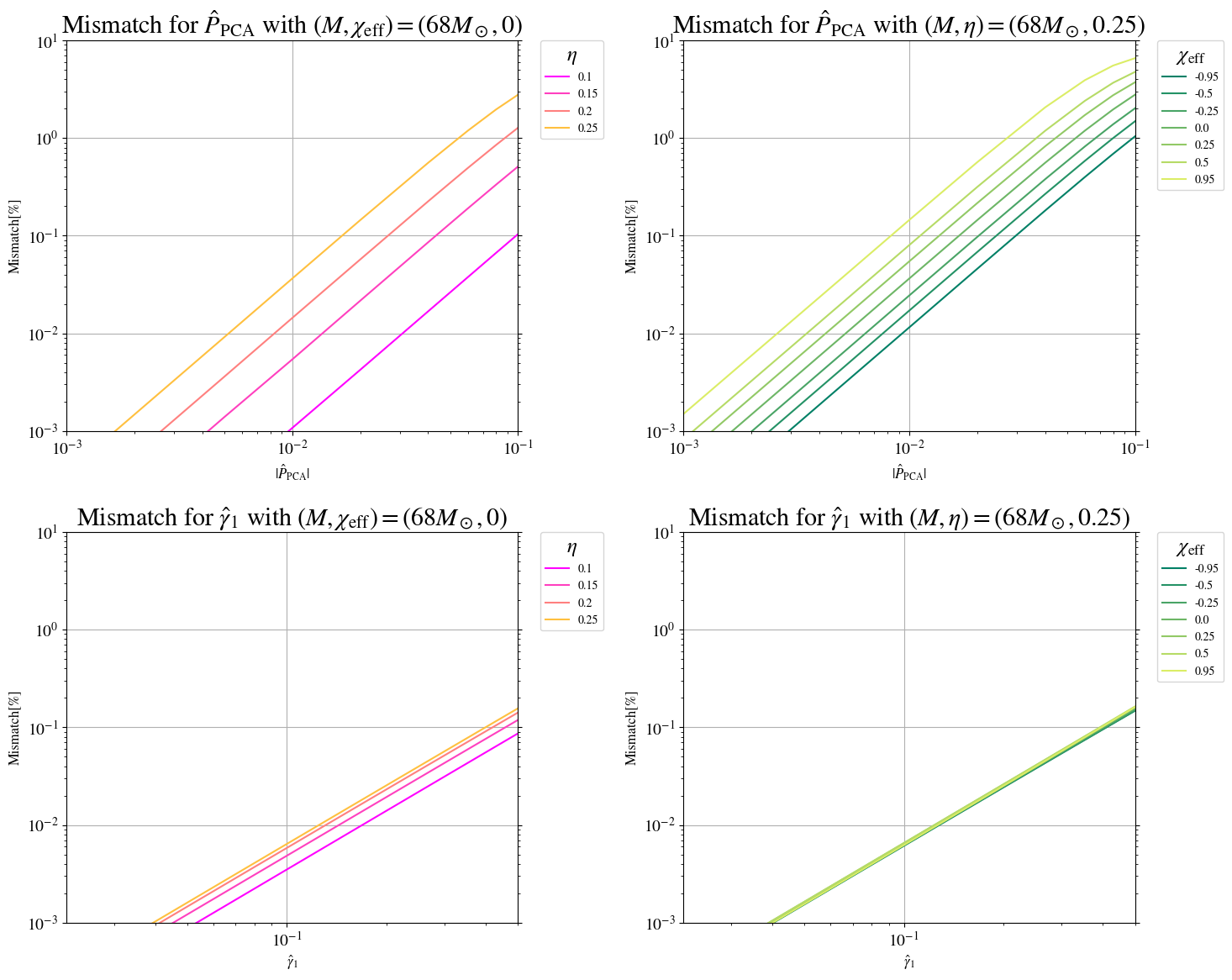}
\caption{\label{fig:mismatches}
$\mathcal{MM}$ associated with $\hat{P}_\mathrm{PCA}$ and $\hat{\gamma}_1$ for various initial configurations. The top two figures show $\mathcal{MM}$ for $\hat{P}_\mathrm{PCA}$, and the left and right show the caseI and caseII, respectively. The bottom two show $\mathcal{MM}$ for $\hat{\gamma}_1$, and the left and right show the caseI and caseII, respectively. For $\hat{P}_\mathrm{PCA}$, $\mathcal{MM}$ is the same for positive and negative values. For $\hat{\gamma}_1$, on the other hand, the values of $\mathcal{MM}$ are not exactly symmetrical for positive and negative values. For a given positive $\hat{\gamma}_1$, $\mathcal{MM}$ for a negative $\hat{\gamma}_1$ with the same absolute value is 1 to 2 times smaller.}
\end{figure*}

\begin{figure*}[t]
\includegraphics[scale=0.27]{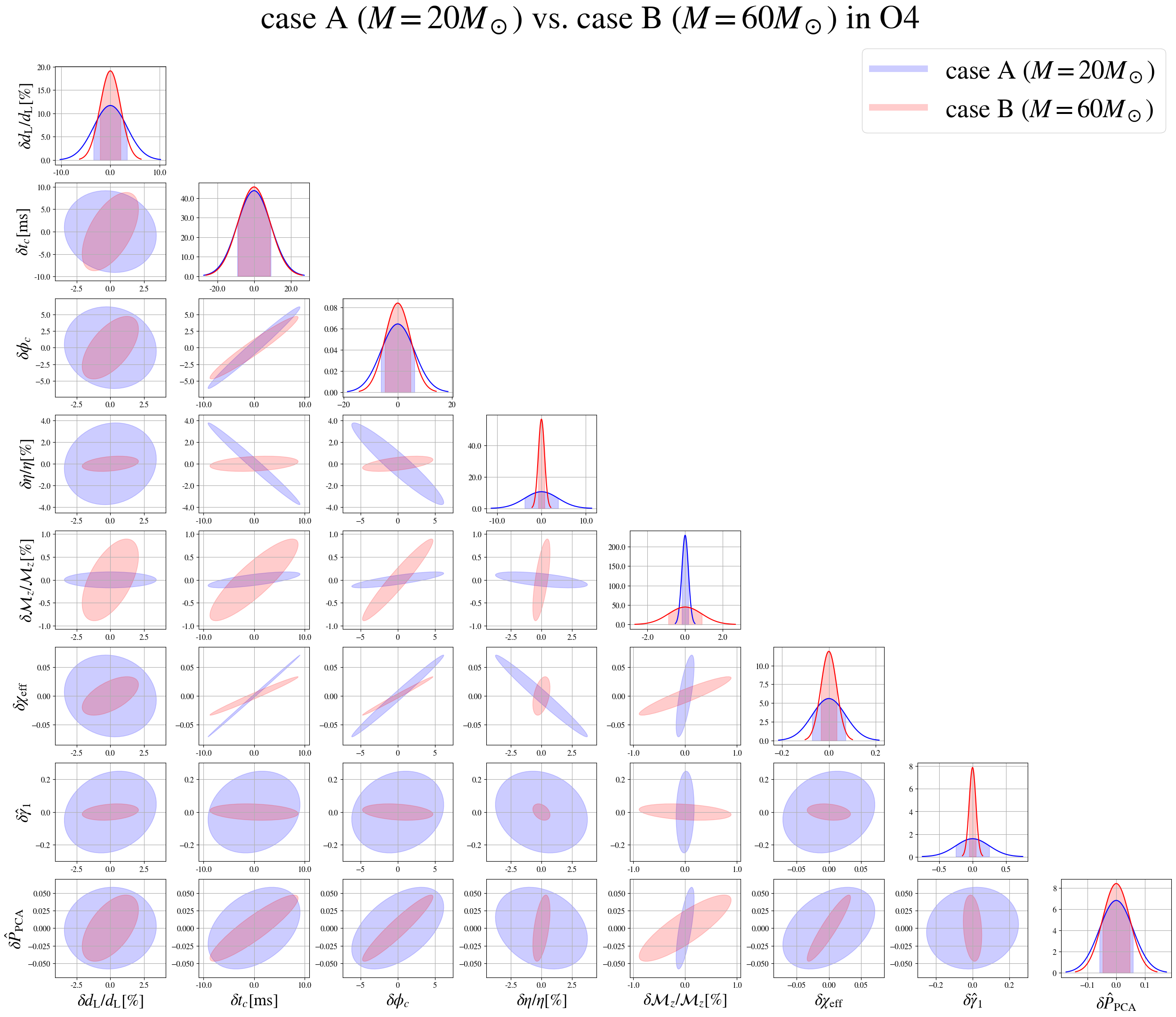}
\caption{\label{fig:Fisher_analysis_the O4}
Marginalized 1$\sigma$ error estimates for $\boldsymbol{\theta}$ in the O4 observation. The blue and the red show the $1\sigma$ regime for case A and case B, respectively.}
\end{figure*}

\begin{figure*}[t]
\includegraphics[scale=0.27]{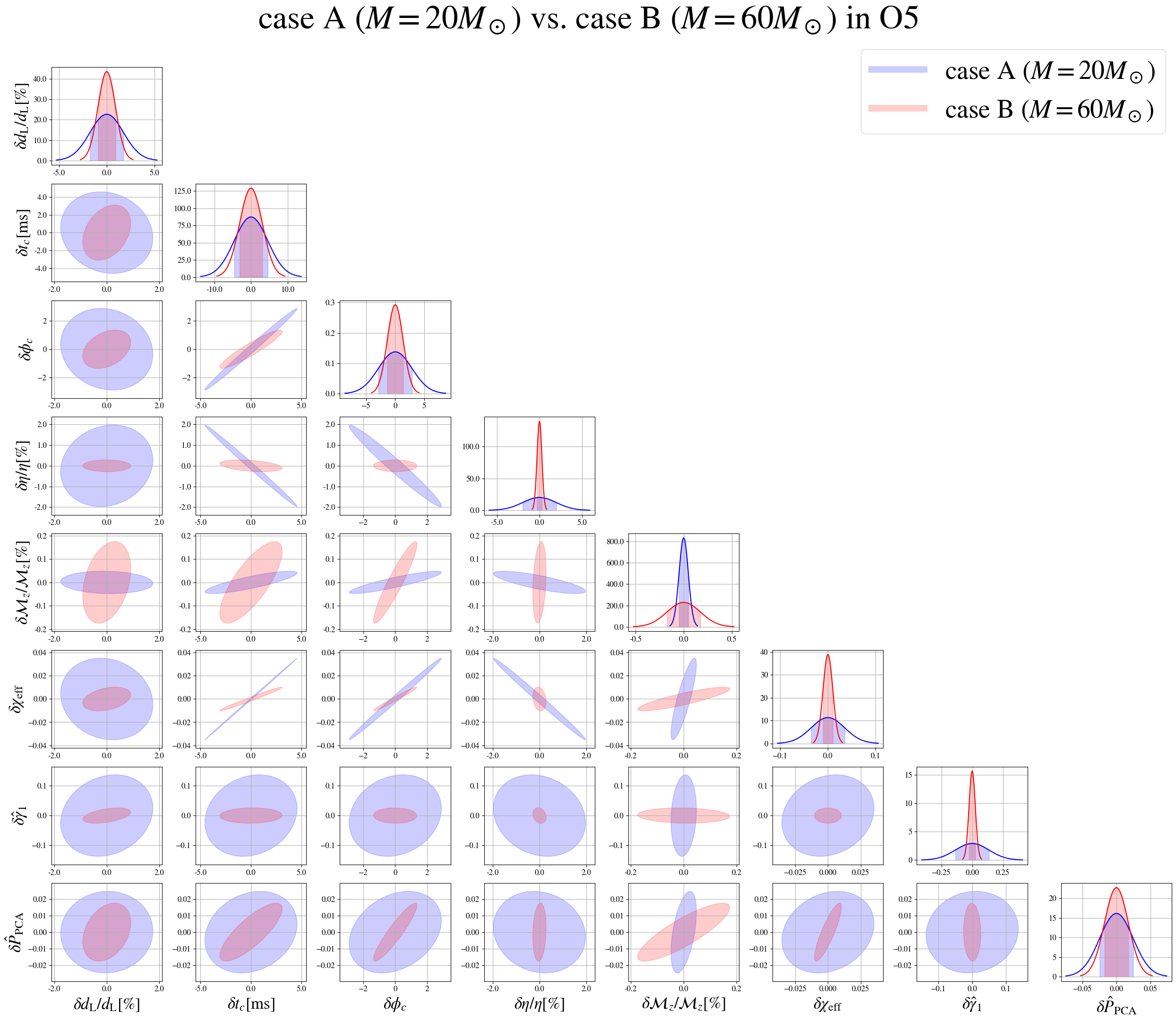}
\caption{\label{fig:Fisher_analysis_O5}
Marginalized 1$\sigma$ error estimates for $\boldsymbol{\theta}$ in the O5 observation. The blue and the red show the $1\sigma$ regime for case A and case B, respectively.}
\end{figure*}

\section{Evaluation of the constructed waveform}
\label{sec:Systematic_studies}

In this section, we discuss the systematic properties of our waveform. In Sec.~\ref{sec:mismatch}, we first study mismatch associated with $\hat{P}_{\mathrm{PCA}}$ and $\hat{\gamma}_1$ for various configurations. The results show what type of BBHs are preferred for testing gravity. In Sec.~\ref{sec:Fisher_analysis}, we estimate the measurement errors and correlation coefficients of the model parameters in the O4 and O5 observations, using the Fisher analysis \cite{Cutler_Flanagan_1994,Poisson_Will_1995}, and show that our parameterization efficiently captures possible deviations from GR waveforms.
Finally, in Sec.~\ref{sec:EdGB}, to show the compatibility with extended theories of gravity, we consider the case of Einstein-dilaton Gauss-Bonnet gravity (EdGB) \cite{EdGB_waveform}, which is a quadratic gravity theory that has an additional scalar-field coupling to a quadratic term of curvature. 

\subsection{Mismatch}
\label{sec:mismatch} 
We adopt a definition of mismatch between the IMRPhenomD waveform, $\tilde{h}_\mathrm{GR}(f)$, in Eq.~(\ref{IMRPhenomD}), and our modified waveform, $\tilde{{h}}_\mathrm{MG}(f)$, in Eq.~\eqref{modified_waveform_final}:
\begin{equation}
\label{mismatch}
\begin{split}
    &\mathcal{MM}\left(\tilde{h}_\mathrm{GR}(f), \tilde{{h}}_\mathrm{MG}(f)\right)\\ &:= 1-\max_{\{t_c,\phi_c\}} \frac{\left(\tilde{h}_\mathrm{GR}(f), \tilde{{h}}_\mathrm{MG}(f)\right)}{\sqrt{\left(\tilde{h}_\mathrm{GR}(f), \tilde{h}_\mathrm{GR}(f)\right)}\sqrt{\left(\tilde{{h}}_\mathrm{MG}(f), \tilde{{h}}_\mathrm{MG}(f)\right)}}\;.
\end{split}
\end{equation}
We investigate $\mathcal{MM}$ associated with $\hat{P}_\mathrm{PCA}$ and $\hat{\gamma}_1$ for configurations with varying $\eta$ and $\chi_\mathrm{eff}$, respectively. Here, we set $S_n(f)=1$ as in Sec.~\ref{sec:phase_modification}, and the lower cutoff is set to $0.0035/M$, which is sufficiently lower than $f=f_\mathrm{p1}$ where deviations occur.
We check for two types of initial setups:
\begin{itemize}
    \item{case I (varying $\eta$)}\mbox{}\\
    $\eta=\{0.1,0.15,0.2,0.25\},  \chi_\mathrm{eff} = 0\;,$
    \item{case II (varying $\chi_\mathrm{eff}$)}\mbox{}\\
    $\eta = 0.25, \\ \chi_\mathrm{eff}=\{-0.95,-0.5,-0.25,0,0.25,0.5,0.95\}\;.$
\end{itemize}
Here $\mathcal{MM}$ for varying $M$ is not considered because of the universality of total-mass scaling in the waveform. 

Figure~\ref{fig:mismatches} shows the results. For case I, $\mathcal{MM}$ tends to be larger as $\eta$ increases for both of $\hat{P}_\mathrm{PCA}$ and $\hat{\gamma}_1$. Similarly, for case II, $\mathcal{MM}$ tends to be larger as $\chi_\mathrm{eff}$ increases for both.
Furthermore, for the case of $\hat{P}_\mathrm{PCA}$, varying $\eta$ or $\chi_\mathrm{eff}$ changes $\mathcal{MM}$ by up to a factor of 10. Therefore, BBH with symmetric mass ratio and positive $\chi_\mathrm{eff}$ is preferred as a best target for testing GR in the nonlinear regime with our waveform. This fact holds for the case of $\hat{\gamma}_1$, though the change is more suppressed than the case of $\hat{P}_\mathrm{PCA}$.

As indicators for the distinguishability of beyond-GR effects, we estimate the minimum SNR values for $\mathcal{MM}$ to be detected. We adopt the criterion proposed in Appendix~G of \cite{katarina_2017},
\begin{equation}
\label{mismatch_criterion}
    \mathcal{MM} \lesssim \frac{D}{2\:\mathrm{SNR}^2}\;,
\end{equation}
where $D$ is the number of the parameters used in the analysis. For our waveform, the parameters are $\{ M, \eta, \chi_\mathrm{eff}, \hat{\gamma}_1, \hat{P}_\mathrm{PCA} \}$, then $D=5$. Following Eq.~\eqref{mismatch_criterion}, for $\mathcal{MM}$ of $1\%$, $0.1\%$, and $0.01\%$, the SNR necessary for detection is $16$, $50$, and $160$, respectively.

\subsection{The Fisher forecast}
\label{sec:Fisher_analysis}
We estimate the statistical errors of beyond-GR parameters for a high-SNR event in the O4 and O5 observations, based on the Fisher matrix formalism 
(e.g.,~\cite{Cutler_Flanagan_1994, Poisson_Will_1995}). Generally, the Fisher analysis is useful in terms of not only error estimation prior to observations but also the assessment of the parameterization of a waveform, that is, investigating the existence of parameter degeneracies.

\subsubsection{\textbf{The Fisher matrix formalism
}}
In the analysis, we consider 8 parameters:
\begin{equation}
    \boldsymbol{\theta} := \{ \log{d_\mathrm{L}}, t_c, \phi_c, \log{\eta}, \log{\mathcal{M}_z}, \chi_\mathrm{eff}, \hat{\gamma}_1, \hat{P}_\mathrm{PCA} \}
\end{equation}
where $d_\mathrm{L}$ is luminosity distance and $\mathcal{M}_z := (1+z)\eta^{\frac{3}{5}}M$ is redshifted chirp mass. As a spin parameter, we focus only on the effective spin parameter, $\chi_\mathrm{eff}$, because it is estimated easily as a leading effect of spins.
The Fisher matrix for each detector labeled by $i=\{$LIGO Hanford, LIGO Livingston, Virgo, KAGRA$\}$ is given as
\begin{equation}
    \Gamma^{(i)}_{ab} := \left( \partial_a \tilde{{h}}_\mathrm{MG}(f), \partial_b \tilde{{h}}_\mathrm{MG}(f) \right)
\end{equation}
where $\partial_a$ stands for the partial derivative with respect to $\theta_a$, and  $S^{(i)}_\mathrm{n}(f)$ in the inner product defined in Eq.~\eqref{inner_product} is the full-sky angular averaged noise spectral density for the $i$-th detector in the O4 and O5 observations~\cite{LIGO2022}. The lower cutoff $f_\mathrm{min}$ is set to 10\,Hz, which is the minimum frequency of the detectors. Given the Fisher matrix for each detector, the statistical error of a parameter in a multi-detector observation, denoted by $\delta \theta_a$, and the correlation coefficients between different parameters, denoted by $c_{\theta_a \theta_b}$, are estimated by
\begin{gather}
    \delta \theta_a = \sqrt{(\Gamma^{-1})_{aa} }, \\
    c_{\theta_a \theta_b} = \frac{(\Gamma^{-1})_{ab}}{\sqrt{(\Gamma^{-1})_{aa} (\Gamma^{-1})_{bb} }},
\end{gather}
where $\Gamma:=\sum\nolimits_{i} \Gamma^{(i)}$ is the combined Fisher matrix and $\Gamma^{-1}$ is the inverse of $\Gamma$.
Fixing source distance to $d_\mathrm{L}=424.6\,\mathrm{Mpc}$ or $z=0.09$ that is consistent with GW150914~\cite{first_detection_gw150914}, mass ratio to $\eta=0.25$, and a leading spin to $\chi_\mathrm{eff} = 0$, we examine two cases of different total masses: $M=\{20M_\odot, 60M_\odot\}$ or $\mathcal{M}_z=\{ 9.5M_\odot, 28.5M_\odot \}$, where $d_\mathrm{L}$ is set to the presented values considering cosmological parameters shown in \cite{Planck_2020}. In the following, we refer to the case of $M=20M_\odot$ and $M=60M_\odot$ as case A and case B, respectively.

\begin{table}[b]
\caption{\label{tb:SNRs}%
SNR value for each parameter set }
\begin{ruledtabular}
\begin{tabular}{ccc}
 case &  the O4 network SNR & O5 network SNR \\
\colrule
case A ($M=20M_\odot$) & 30.4 & 58.9 \\
case B ($M=60M_\odot$) & 70.2 & 136.3 \\
\end{tabular}
\end{ruledtabular}
\end{table}

\begin{table*}[t]
\begin{ruledtabular}
\caption{\label{tb:errors}%
1$\sigma$ errors of $\theta$}
\begin{tabular}{ccccccccc}
 case & $\delta d_\mathrm{L}/d_\mathrm{L} [\%]$ & $t_c [\mathrm{ms}]$ & $\phi_c$ & $\delta \eta/\eta [\%]$ & $\delta \mathcal{M}_z/\mathcal{M}_z [\%]$ & $\delta \chi_\mathrm{eff}$ & $\delta \hat{\gamma}_1$ & $\delta \hat{P}_\mathrm{PCA}$ \\
\colrule
A in O4 & 3.41 & 9.12 & 6.17 & 3.77 & 0.174 & 0.0712 & 0.251 & 0.0583 \\
A in O5 & 1.76 & 4.57 & 2.89 & 1.97 & 0.0480 & 0.0352 & 0.136 & 0.0247 \\
B in O4 & 2.09 & 8.74 & 4.73 & 0.706 & 0.889 & 0.0335 & 0.0505 & 0.0473 \\
B in O5 & 0.914 & 3.09 & 1.36 & 0.285 & 0.174 & 0.0102 & 0.0255 & 0.0176\\
\end{tabular}
\end{ruledtabular}
\end{table*}
\begin{table*}[t]
\caption{\label{tb:correlation_gamma1}%
Correlation coefficients of $\hat{\gamma}_1$}
\begin{ruledtabular}
\begin{tabular}{cccccccc}
 case & $c_{ \hat{\gamma}_1 \log{d_\mathrm{L}}}$ & $c_{ \hat{\gamma}_1 t_c}$ & $c_{ \hat{\gamma}_1 \phi_c}$ & $c_{ \hat{\gamma}_1 \eta}$ & $c_{ \hat{\gamma}_1 \log{M}_z}$ & $c_{ \hat{\gamma}_1 \chi_\mathrm{eff}}$ & $c_{ \hat{\gamma}_1 \hat{P}_\mathrm{PCA} }$ \\
\colrule
A in O4 & 0.199 & 0.183 & 0.176 & -0.184 & 0.110 & 0.182 & 0.0941 \\
A in O5 & 0.186 & 0.174 & 0.173 & -0.174 & 0.140 & 0.174 & 0.0799 \\
B in O4 & 0.246 & -0.181 & -0.258 & -0.348 & -0.307 & -0.230 & -0.220 \\
B in O5 & 0.460 & 0.0449 & -0.0355 & -0.212 & -0.102 & 0.00520 & -0.0363 \\
\end{tabular}
\end{ruledtabular}
\end{table*}
\begin{table*}[t]
\caption{\label{tb:correlation_P_PCA}%
Correlation coefficients of $\hat{P}_\mathrm{PCA}$}
\begin{ruledtabular}
\begin{tabular}{ccccccc}
 case & $c_{ \hat{P}_\mathrm{PCA} \log{d_\mathrm{L}}}$ & $c_{ \hat{P}_\mathrm{PCA} t_c}$ & $c_{ \hat{P}_\mathrm{PCA} \phi_c}$ & $c_{ \hat{P}_\mathrm{PCA} \eta}$ & $c_{ \hat{P}_\mathrm{PCA} \log{M}_z}$ & $c_{ \hat{P}_\mathrm{PCA} \chi_\mathrm{eff}}$ \\
\colrule
A in O4 & 0.0448 & 0.391 & 0.514 & -0.263 & 0.722 & 0.410 \\
A in O5 & 0.0272 & 0.203 & 0.288 & -0.121 & 0.498 & 0.206 \\
B in O4 & 0.550 & 0.904 & 0.942 & 0.516 & 0.845 & 0.924 \\
B in O5 & 0.326 & 0.804 & 0.922 & 0.142 & 0.846 & 0.863 \\
\end{tabular}
\end{ruledtabular}
\end{table*}

\subsubsection{\textbf{Results}}
Table~\ref{tb:SNRs} shows the network SNR values with our modified waveform for each configuration when detected in the O4 and O5 observations. 
In Figs~\ref{fig:Fisher_analysis_the O4} and \ref{fig:Fisher_analysis_O5}, each panel shows the 1$\sigma$ error ellipse marginalized over other parameters 1$\sigma$ errors for the O4 and O5 observations, respectively. 
The numerical values of the errors and correlation coefficients for $\hat{\gamma}_1$ and $\hat{P}_\mathrm{PCA}$ are shown in Tables~\ref{tb:errors}, \ref{tb:correlation_gamma1}, and \ref{tb:correlation_P_PCA}, respectively.

For the GR parameters, Table~\ref{tb:errors} shows that the estimated error of $\mathcal{M}_z$ is smaller in case A than in case B, contrary to the errors of the other GR parameters, for both the O4 and O5 observations. This is because the inspiral signal of a binary with small total mass lasts longer up to a larger orbital frequency, then detectors can see the early inspiral regime more clearly than a binary with large total mass. Due to this, for case A, degeneracy between $\mathcal{M}_z$ and the other GR parameters are solved, however, correlations among $t_c$, $\phi_c$, $\eta$, and $\chi_\mathrm{eff}$ are strong. On the other hand, for case B, we can see the effects of higher-order PN coefficients more efficiently, thus correlations among $t_c, \phi_c, \eta$, and $\chi_\mathrm{eff}$ are solved more clearly than for case A, as  Figs.~\ref{fig:Fisher_analysis_the O4} and \ref{fig:Fisher_analysis_O5} show. 

According to Tables~\ref{tb:correlation_gamma1} and \ref{tb:correlation_P_PCA}, although $\hat{P}_{\rm PCA}$, and $\phi_{\rm c}$ or $t_{\rm c}$ are correlated modestly, the correlations between the GR parameters and beyond-GR parameters are not significant. This fact means that we can expect to estimate beyond-GR effects easily by analyzing the data with our waveform.

\subsection{Compatibility with EdGB waveform}
\label{sec:EdGB}
Although our waveform does not depend on specific models of gravity theories, we should investigate what theories are compatible with the waveform.
Recently numerical simulations in extended gravity theories have been developed and produced IMR waveforms in such theories~\cite{EdGB_waveform, dCS_waveform}.
Here, to see compatibility, we take Einstein-dilation Gauss-Bonnet(EdGB) gravity waveforms~\cite{EdGB_waveform}. EdGB gravity is an effective field theory that has a scalar field coupling to a specific combination of quadratic curvature terms, so-called the Gauss-Bonnet term~\cite{Fernandes_2022}. The action is written as
\begin{equation}
S := \int \frac{m^2_\mathrm{pl}}{2} \diff^4 x \sqrt{-g} \left[ R - \frac{1}{2}(\partial \theta)^2 + 2 \alpha_\mathrm{GB} f(\theta) \mathcal{R}_\mathrm{GB}  \right]\;,
\end{equation}
where $m_\mathrm{pl} $ is the reduced Planck mass, $R$ is the 4-dimensional Ricci scalar, $\theta$ is the EdGB scalar field, $\alpha_\mathrm{GB}$ is the EdGB coupling constant with dimensions of length squared, $\mathcal{R}_\mathrm{GB}$ is the EdGB scalar,
\begin{equation}
    \mathcal{R}_\mathrm{GB} = R^{abcd}R_{abcd} - 4R^{ab}R_{ab} + R^2\;,
\end{equation}
and the function, $f(\theta)$, takes a form of $f(\theta) = \frac{1}{8}\mathrm{e}^{\theta}$ in this theory (e.g.,~\cite{Helvi_2019}). 

Although it is unknown whether EdGB gravity has well-posedness for the initial value problem, the waveforms are computed in \cite{EdGB_waveform} based on an order-reduced scheme in which well-posedness of the EOM is ensured at each order of perturbations. The method is as follows. Given the NR data of a BBH merger as the background spacetime, the background sources the leading-order scalar field. Then, the GR background and leading-order scalar field source the leading-order correction on a gravitational waveform. Here, as the background spacetime, GR waveforms computed by SXS colaboration~\cite{Boyle_2019} are adopted. To obtain EdGB waveforms, we use a Python code provided by Okounkova~\cite{EdGB_web}.
Since the backreaction due to the additional GWs is not considered in the EdGB waveforms, we turn off the backreaction inclusion in the fitting. 

\begin{table}[b]
\caption{\label{tb:RSS}%
$\mathrm{RSS}/\mathrm{RSS}_\mathrm{GR}$ for $\sqrt{\alpha_\mathrm{GB}}/GM=\{0.030,0.035,0.040\}$}
\begin{ruledtabular}
\begin{tabular}{ccccc}
$\sqrt{\alpha_\mathrm{GR}}/GM$ & $0$(GR) & $0.030$ & $0.035$ & $0.040$ \\ \hline
    $\mathrm{RSS}/\mathrm{RSS}_\mathrm{GR}$ & 1 & 2.30\: & 3.33\: & 7.45\: \\
\end{tabular}
\end{ruledtabular}
\end{table} 

\begin{figure*}[t]
\includegraphics[scale=0.39]{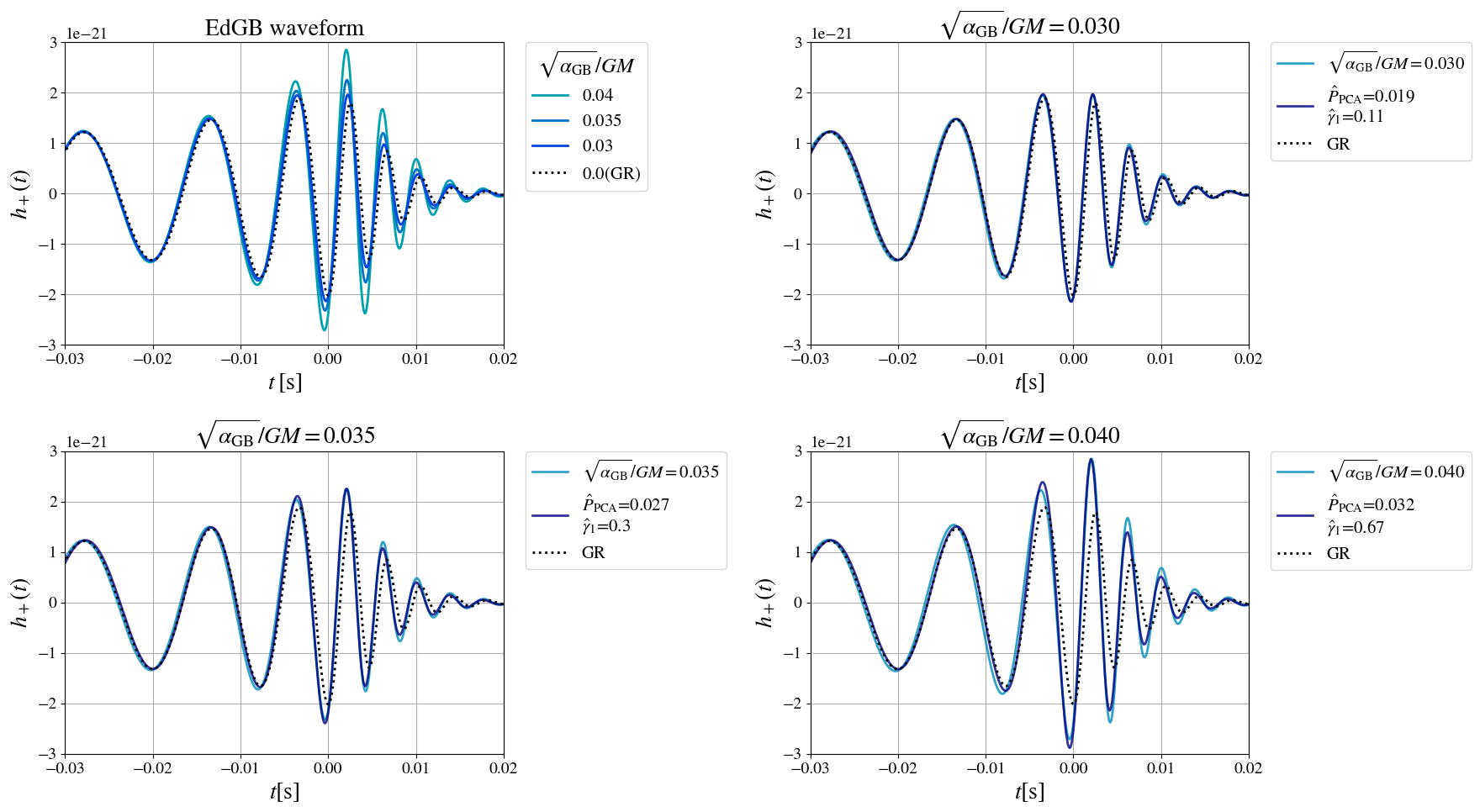}
\caption{\label{fig:EdGB}
Best-fit parameterized waveforms to EdGB waveforms. The top left panel shows EdGB waveforms for $\sqrt{\alpha_\mathrm{GB}}/GM=0.030, 0.035, 0.040$ and GR waveform (dotted). From the top right to the bottom right, the best-fit modified waveforms (light blue lines) are shown for $\sqrt{\alpha_\mathrm{GB}}/GM=0.030, 0.035, 0.040$ (blue lines).} 
\end{figure*}

\subsubsection{\textbf{Fitting results}}
We evaluate the residual sum of squared (RSS) between an EdGB waveform and the best-fit modified waveform for the goodness of fit. Note that there is a systematic error in the modeling of the IMRPhenomD waveform. Thus, we consider a fraction, $\mathrm{RSS}/\mathrm{RSS}_\mathrm{GR}$, where $\mathrm{RSS}_\mathrm{GR}$ is the RSS value computed from the IMRPhenomD and SXS waveforms~\cite{Boyle_2019}. In this analysis, we use the GW150914-like parameters in Table~\ref{tb:binary}.

We fit the cases of $\sqrt{\alpha_\mathrm{GB}}/GM=\{ 0.030, 0.035, 0.040 \}$ here. These coupling constants are chosen not to drastically exceed the perturbative regime of $|\hat{\gamma}_1|$. The results are presented in Table~\ref{tb:RSS}. Figure~\ref{fig:EdGB} shows the best-fit modified waveform in the time domain for each case.

\subsubsection{\textbf{Discussion}}
Note that, unlike our original waveform, EdGB
waveforms computed in \cite{EdGB_waveform} do not include backreactions to the background spacetime. In other words, although the EdGB waveform has a deformation from GR in the ringdown part, it is purely caused by the scalar field. In contrast, our waveform includes the backreactions due to additional gravitational radiation. Therefore, we cannot expect our waveform to be fully compatible, though we turn off the inclusion of the backreaction here. Regardless of the small difference in modeling, the compatibility should be evaluated by comparing the goodness of fit with the parameter estimation errors in future observations.

As a benchmark to whether our waveform can reproduce EdGB waveforms, we use $\mathrm{RSS}_\mathrm{O4,5}$, which is defined as the RSS value computed from the IMRPhenomD waveform with binary parameters shown in Table~\ref{tb:binary} (i.e, $\hat{P}_\mathrm{PCA}=\hat{\gamma}_1=0$) and the modified waveform with setting beyond-GR parameters to $1\sigma$ errors in the O4 or O5 observations, that is, $\hat{P}_\mathrm{PCA} \rightarrow \delta\hat{P}_\mathrm{PCA}$ and $ \hat{\gamma}_1 \rightarrow \delta\hat{\gamma}_1$ with binary parameters set to values shown in Table~\ref{tb:binary}. Note that a degeneracy between $\hat{P}_\mathrm{PCA}$ and $\hat{\gamma}_1$ turns out to be insignificant as shown in Sec.~\ref{sec:Fisher_analysis}. $\mathrm{RSS}_\mathrm{O4,5}$ are arose from practical statistical errors, which are inevitable in the parameter estimation of real data. For this parameter set, $\mathrm{RSS}_\mathrm{O4}/\mathrm{RSS}_\mathrm{GR} = 19.9$ and $\mathrm{RSS}_\mathrm{O5}/\mathrm{RSS}_\mathrm{GR} = 4.06$.

According to Table~\ref{tb:RSS}, all cases are well below or comparable to $\mathrm{RSS}_\mathrm{O4,5}$. Therefore, we can conclude that, in both the O4 and O5 observations, our waveform model is compatible with EdGB waveforms in the perturbative ranges of $\hat{P}_\mathrm{PCA}$ and $\hat{\gamma}_1$. According to Figure~\ref{fig:EdGB}, in terms of compatibility with EdGB waveforms, our waveforms tend to fit larger in the pre-merger portion and smaller in the ringdown regime than EdGB waveforms. Another important conclusion is that, for phase, it is sufficient to introduce one additional parameter to fit EdGB waveforms. This fact supports that a leading principal component plays a crucial role in interpreting physics behind deviations from GR.

\section{Discussion}
\label{sec:Discussion}
There are several factors to consider for the availability of our physically consistent waveform. Here, we give some comments on these issues as discussions.
\begin{itemize}
    \item {\textbf{Energy and angular momentum carried by the nonstandard effects}}\\
    For our waveform to describe the physically consistent situation, we assume the amounts of energy and angular momentum losses carried or consumed by the nonstandard effects, such as an additional field around the source or environmental effects around BHs, are negligible compared with the ones carried by GWs. If this assumption breaks, we will find a deviation from GR in the estimated results using our waveform, especially in the ringdown part, even if our waveform captures such nonstandard effects as deviations from GR with non-zero $\hat{\gamma}_1$. The estimate of this deviation is expected to allow us to infer the nature of the additional physical degrees of freedom, such as non-tensorial GW modes.
    
    \item {\textbf{Energy and angular momentum carried by the sub-dominant GW modes}}\\
    In this study, we also assume the amounts of energy and angular momentum carried by the sub-dominant modes, such as $(l,m)=(3,\pm 3)$ or $(l,m)=(2,\pm 1)$ modes, can be neglected. This is just because we use the IMRPhenomD model, which describes $(l,m)=(2, \pm 2)$ modes as a basis for the first implementation. Indeed, if we focus on BBHs with nearly equal component masses, this assumption is reasonable if the event SNR is not so high, or for the events detected so far. However, for a high SNR event, sub-dominant modes should be considered for a precise analysis because this simplification can cause confusion with beyond-GR effects. 
    
    \item {\textbf{Precession effects}}\\
    Since we adopt the IMRPhenomD waveform as a basis, we do not consider the precession effects of BBHs. This effect also can be confused with beyond-GR effects, especially in the analysis of a high SNR event. Therefore, the method that we have developed in this paper should be implemented to the latest waveforms including the sub-dominant modes or, and precession effects, such as \cite{garcia_2020, Geraint_2021}.
\end{itemize}
Regarding the second and third items, our idea, whose main features are to consider the principal components of the merger parameters and to include the radiation reaction due to beyond-GR parameters, is in principle applicable to the latest waveforms. From this point, there is no obstacle to the implementation. However, technical difficulties might arise since the latest waveforms have a larger number of phenomenological parameters. We should reconsider an appropriate criterion to reduce the number of beyond-GR parameters in a modified waveform effectively.

\section{Conclusions}
\label{sec:Conclusion}
In this work, we propose a parameterized waveform that can measure generic deviations from GR in the non-linear regime. Our underlying idea is that perturbative modifications to a GR waveform can capture beyond-GR effects in the nonlinear regime of gravity. As the first implementation, we use the IMRPhenomD waveform~\cite{IMRPhenomD_2}. For modification to GW phase, we adopt only the leading principal component of the artificial parameters as a beyond-GR parameter, $\hat{P}_\mathrm{PCA}$, because the leading component is more significant than the others, as discussed in Sec.~\ref{sec:specification_PCA}. Our ideas here are mainly two. The first is that the dominant components are expected to play physically crucial roles, though it is difficult to interpret the physical meanings of the artificial parameters. The other is to break the degeneracies between the estimated parameters that have not been considered in the previous studies (e.g.,~\cite{testing_GR_GW150914,testing_GR_GWTC1,testing_GR_GWTC2}).
$\hat{P}_\mathrm{PCA}$ corresponds to a mixture of time and phase shifts to a GR case. For modification to GW amplitude, we adopt the fractional change of $\gamma_1$, $\hat{\gamma}_1$, which controls amplitude around the peak of a waveform. As a role of $\hat{\gamma}_1$, we interpolate the inspiral part linearly to the midway frequency such that $\hat{\gamma}_1$ also describes an amplification in the intermediate regime. The key point is that our modified waveform is physically consistent in the sense that the radiation backreactions associated with $\hat{\gamma}_1$ are included. This inclusion is achieved by calculating the additional gravitational radiation and modifying mass and spin of a remnant BH. Thus, the ringdown signal is modified. 

Furthermore, we investigate the systematic studies for the modified waveform, specifically considering mismatch associated with the beyond-GR parameters, the Fisher analysis, and the compatibility with EdGB waveforms. In Sec.~\ref{sec:mismatch}, we find that mismatch can be more significant for a binary that has a symmetric mass ratio and positive $\chi_\mathrm{eff}$ value when fixing total mass and luminosity distance. Therefore, we conclude that such binaries are preferred for our analysis. In Sec.~\ref{sec:Fisher_analysis}, we estimate the expected error of each parameter and the correlation coefficients between the parameters. GWs from BBHs with a small total mass allow us to derive the posterior of $\mathcal{M}_z$ more tightly than with a large total mass, due to the long inspiral. For other parameters, contrary to $\mathcal{M}_z$, the errors are smaller when considering GWs come from BBHs with a large total mass. The important fact for our waveform parameterization is that newly introduced parameters, $\hat{\gamma}_1$ and $\hat{P}_\mathrm{PCA}$, are not correlated with others significantly. This means that we can expect beyond-GR parameters will be estimated independently from GR parameters. Finally, we have shown that our modified waveform can reproduce EdGB waveforms in perturbative ranges of $\hat{P}_\mathrm{PCA}$ and $\hat{\gamma}_1$, within the measurement errors in the O4 and O5 observations. It is remarkable that only two additional parameters are sufficient to describe an extended theory. This fact supports that the principal components play a crucial role in interpreting physics behind deviations from GR. 

\begin{acknowledgments}
We thank Hiroyuki Nakano and Hiroki Takeda for useful discussions.
D. W. is supported by JSPS KAKENHI grant No. 23KJ06945. A. N. was supported by JSPS KAKENHI Grant No. JP20H04726 and Research Grants from Inamori Foundation in the initial stage of the research and is supported by JSPS KAKENHI Grant Nos. JP23K03408, JP23H00110, and JP23H04893.
\end{acknowledgments}

\appendix

\section{Fitting formula of $\Vec{e}_\mathrm{PCA}$}
\label{sec:fitting formula}
\begin{figure}[t]
\includegraphics[scale=0.45]{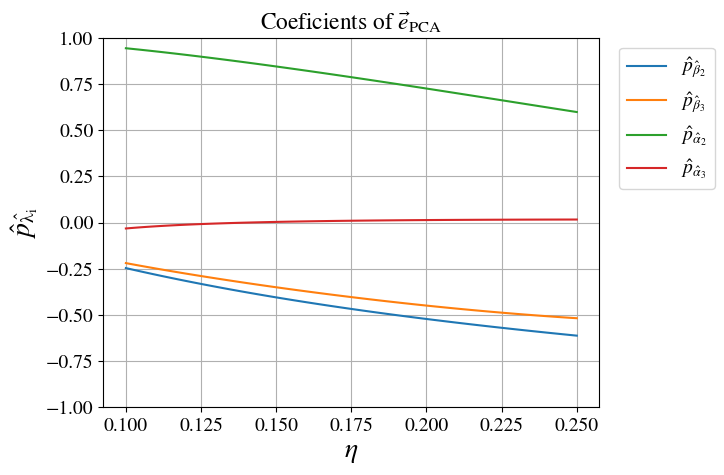}
\caption{\label{fig:coefficients_fitting}
Coefficients of $\vec{e}_\mathrm{PCA}$ with $0.1\leq\eta\leq0.25$ and $\chi_\mathrm{eff}=0$. Each line shows $\hat{p}_\mathrm{\hat{\beta}_2}$ (blue), $\hat{p}_\mathrm{\hat{\beta}_3}$ (orange) $\hat{p}_\mathrm{\hat{\alpha}_2}$ (green) and $\hat{p}_\mathrm{\hat{\alpha}_3}$ (red).} 
\end{figure}

$\vec{e}_\mathrm{PCA}$ has a form of
\begin{equation}
    \vec{e}_\mathrm{PCA} = \hat{p}_{\hat{\beta}_2} \vec{e}_{\hat{\beta}_2} + \hat{p}_{\hat{\beta}_3} \vec{e}_{\hat{\beta}_3} +
    \hat{p}_{\hat{\alpha}_2} \vec{e}_{\hat{\alpha}_2} +
    \hat{p}_{\hat{\alpha}_3} \vec{e}_{\hat{\alpha}_3}\;.
\end{equation}
Figure~\ref{fig:coefficients_fitting} shows the coefficients of $\vec{e}_\mathrm{PCA}$ with $0.1 \leq \eta \leq 0.25$ fixing $\chi_\mathrm{eff}=0
$.
Here, for $0.1 \leq \eta \leq 0.25$ and $-0.95 \leq \chi_\mathrm{eff} \leq 0.95$,  we give fitting formulas for the coefficients. We adopt a polynomial ansatz using $(\eta, \chi_\mathrm{eff})$,
\begin{multline}
    \hat{p}_{\hat{\lambda}_i} =  \hat{p}^i_{00} + \hat{p}^i_{10} \eta + (\chi_\mathrm{eff}-1)(\hat{p}^i_{01} + \hat{p}^i_{11}\eta + \hat{p}^i_{21}\eta^2) \\
    + (\chi_\mathrm{eff}-1)^2(\hat{p}^i_{02} + \hat{p}^i_{12}\eta + \hat{p}^i_{22}\eta^2)\\
    +(\chi_\mathrm{eff}-1)^3(\hat{p}^i_{03} + \hat{p}^i_{13}\eta + \hat{p}^i_{23}\eta^2)\;,
\end{multline}
which is motivated by Eq.~~(31) in \cite{IMRPhenomD_2}. In this study, we adopt rather $\chi_\mathrm{eff}$ instead of $\chi_\mathrm{PN}$ used in \cite{IMRPhenomD_2}. The results are shown in Table~\ref{tab:fittingCoefficients}. The errors of $\hat{p}_{{\beta}_2}, \hat{p}_{{\beta}_3}$ and $\hat{p}_{{\alpha}_2}$ in the fitting are below $0.002\%$. On the other hand, the error of $\hat{p}_{{\alpha}_3}$, which has the smallest contribution to $\vec{e}_\mathrm{PCA}$, is below $1\%$.

\begin{table}[t]
\centering
\begin{ruledtabular}
\begin{tabular}{ccccc}
 & $\hat{\beta}_2$ & $\hat{\beta}_3$ & $\hat{\alpha}_2$ & $\hat{\alpha}_3$ \\
 \hline
$\hat{p}^i_{00}$ & -0.291715 & -0.661106 & 1.13042 & 0.189428 \\
$\hat{p}^i_{10}$ & -1.38872 & 0.347803 & -2.39817 & -0.699968 \\
$\hat{p}^i_{01}$ & -0.796214 & -2.33506 & 0.553463 & 0.729275 \\
$\hat{p}^i_{11}$ & 4.62917 & 11.4896 & -7.70244 & -3.52067 \\
$\hat{p}^i_{21}$ & -9.50827 & -8.47591 & 19.4324 & 1.78907 \\
$\hat{p}^i_{02}$ & 0.163208 & -2.21242 & 1.06718 & 0.492195 \\
$\hat{p}^i_{12}$ & -2.71537 & 9.38284 & -11.1916 & -1.29497 \\
$\hat{p}^i_{22}$ & 4.82752 & -1.65164 & 26.416 & -3.77372 \\
$\hat{p}^i_{03}$ & 0.513489 & -0.655432 & 0.526777 & 0.0692508 \\
$\hat{p}^i_{13}$ & -4.00258 & 2.24979 & -4.45295 & 0.310426 \\
$\hat{p}^i_{23}$ & 7.62501 & 0.944151 & 9.58594 & -2.80064 \\
\end{tabular}
\caption{Fitting coefficients for each parameter. These values are derived using Mathematica's NonlinearModelFit.}
\label{tab:fittingCoefficients}
\end{ruledtabular}
\end{table}

\section{Time and phase shifts due to $\hat{P}_\mathrm{PCA}$}
\label{sec:time and phase shifts}
Here we derive quantitative representations of the time and phase shifts associated with $\hat{P}_\mathrm{PCA}$. Since $\Delta\beta_{0,1}, \Delta\alpha_{0,1}$ appear in constants and linear terms for frequency, they can be interpreted as time and phase shifts. First, $\Delta\beta_{0,1}$ induces the shifts at the beginning of the intermediate,
\begin{gather}
    \begin{split}
    \Delta \phi_\mathrm{int} &:= -\frac{\Delta \beta_0}{\eta}\\
    &=-\frac{1}{\eta}\left[ 
\{1-\log(f_\mathrm{p1})\}\Delta\beta_2+\frac{4}{3}{f_\mathrm{p1}}^3\Delta\beta_3 \right]\;,
    \end{split}\\
    \begin{split}
    \Delta t_\mathrm{int} &:= \frac{M}{2\pi\eta}\Delta\beta_1 \\
    &=-\frac{M}{2\pi\eta}\left\{ {f_\mathrm{p1}}^{-1}\Delta\beta_2 + {f_\mathrm{p1}}^{-4}\Delta\beta_3
 \right\}\;,
    \end{split}
 \end{gather}
and, similarly $\Delta \alpha_{0,1}$ induces the shifts at the beginning of the merger\&ringdown,
\begin{gather}
    \begin{split}
    \Delta \phi_\mathrm{MR} &:= -\frac{\Delta\alpha_0}{\eta}\\
        &
        \begin{split}
            =\Delta \phi_\mathrm{int} - \frac{1}{\eta}\Bigl[ &-\{1-\log(f_\mathrm{p2})\}\Delta\beta_2 - \frac{4}{3}{f_\mathrm{p2}}^{-3}\Delta\beta_3\\ &+ 2{f_\mathrm{p2}}^{-1}\Delta\alpha_2 - \frac{1}{3}{f_\mathrm{p2}}^\frac{3}{4}\Delta\alpha_3 \Bigr]\;,
        \end{split} \label{dev_t_MR}
    \end{split}
\end{gather}
\begin{gather}
    \begin{split}
    \Delta t_\mathrm{MR} &:= \frac{M}{2\pi\eta}\Delta\alpha_1 \\
        &
        \begin{split}
             =\Delta t_\mathrm{int} + \frac{M}{2\pi\eta}\Bigl(&{f_\mathrm{p2}}^{-1}\Delta\beta_2 + {f_\mathrm{p2}}^{-4}\Delta\beta_3\\ &- {f_\mathrm{p2}}^{-2}\Delta\alpha_2 - {f_\mathrm{p2}}^{-\frac{1}{4}}\Delta\alpha_3 \Bigr)\;.
        \end{split} \label{dev_phi_MR}\\
    \end{split}
\end{gather}    
The first terms in Eqs.~\eqref{dev_t_MR} and \eqref{dev_phi_MR} are carried over from the inspiral. The second term in each Eq.~ represents the time and phase deviation in the nonlinear regime, respectively. Since homogeneous time and phase shifts are interpreted as GR part when evaluating mismatch (see in Sec.~\ref{sec:mismatch}), these quantities characterize deviations that cannot be absorbed into GR.

\section{Derivation of Eqs.~\eqref{DeltaE} and \eqref{DeltaJz}}
\label{sec:derivation of formulas}
We focus only on $(l,m)=(2,\pm 2)$ components in Eq.~~\eqref{H}, which IMRPhenomD models. Using  Fourier transform of ${h}_{22}(t)$ and ${h}_{2-2}(t)$, $H(t) $ and $\dot{H}(t)$ are written as
\begin{widetext}
\begin{gather}
\label{tilde_H}
H(t) = \int^{\infty}_{-\infty} \diff f \left({}^{-2}Y_{22}(\theta,\phi)\tilde{h}_{22}(f) + {}^{-2}Y_{2-2}(\theta,\phi)\tilde{h}_{2-2}(f) \right) e^{-2\pi i f t}\;, \\
\label{dot_tilde_H}
\dot{H}(t) = \int^{\infty}_{-\infty} \diff f \:(-2\pi i f)\left({}^{-2}Y_{22}(\theta,\phi)\tilde{h}_{22}(f) -{}^{-2}Y_{2-2}(\theta,\phi)\tilde{h}_{2-2}(f) \right) e^{-2\pi i f t}\;.
\end{gather}
\end{widetext}
Here, we have two things to pay attention to.
The first one is that the choice of the Fourier transform is fixed as in Eq.~\eqref{tilde_H}, to ensure that $\tilde{h}_{2m}(f)$ with $m=2$ and $m=-2$ have support for $f>0$ and $f<0$, respectively. The other is that each mode has a different time-to-frequency correspondence in the stationary phase approximation (e.g., \cite{cotesta_2020}), which determines the conversion rule from NR waveform to the frequency-domain waveform in the Phenom modeling. In this case, the relation of this correspondence between $(l=2,m=2)$ mode and $(l=2,m=-2)$ mode is $t^{22}(f)=t^{2-2}(-f)$, therefore, time derivative of $h_{2-2}(t)$ in Eq.~\eqref{dot_tilde_H} must have a minus sign.

Plugging Eq.~\eqref{tilde_H}, \eqref{dot_tilde_H} into Eq.~\eqref{dE} and using the ortho-normality of spin-weighted spherical harmonics, 
\begin{equation}
\int \diff \Omega \left( {}^{s}Y_{lm}(\theta, \phi) {}^{s'}Y^\ast_{l'm'}(\theta, \phi) \right) = \delta^{ss'} \delta_{mm'} \delta_{ll'}\;,
\end{equation}
we derive
\begin{widetext}
\begin{equation}
\begin{split}
    \Delta E &= \frac{c^3 r^2}{16\pi G} \int^{\infty}_{-\infty} \diff t \int \diff \Omega |\dot{H}|^2 \\
    &= \frac{c^3 r^2}{16\pi G} \int^{\infty}_{-\infty} \diff t \int \diff \Omega
    \left[
    \begin{split}
    &\left(\int^{\infty}_{-\infty} \diff f \:(-2\pi i f)\left({}^{-2}Y_{22}(\theta,\phi)\tilde{h}_{22}(f) -{}^{-2}Y_{2-2}(\theta,\phi)\tilde{h}_{2-2}(f) \right) e^{-2\pi i f t}\right) \\
    &\cdot{} \left(\int^{\infty}_{-\infty} \diff f' \:(2\pi i f')\left({}^{-2}Y^\ast_{22}(\theta,\phi)\tilde{h}^\ast_{22}(f') -{}^{-2}Y^\ast_{2-2}(\theta,\phi)\tilde{h}^\ast_{2-2}(f') \right) e^{2\pi i f' t}\right)
    \end{split} \right] \\
    &= \frac{\pi c^3 r^2}{4 G} \iint^{\infty}_{-\infty} \diff f \diff f' \:ff' \left( \tilde{h}_{22}(f)\tilde{h}^\ast_{22}(f) + \tilde{h}^\ast_{22}(f')\tilde{h}_{22}(f') \right) \int^{\infty}_{-\infty} \diff t \:e^{2\pi i (f-f')t} \\
    &= \frac{\pi c^3 r^2}{4 G} \int^{\infty}_{-\infty} \diff f \:f^2\left(\tilde{h}_{22}(f)\tilde{h}^\ast_{22}(f) + \tilde{h}_{2-2}(f)\tilde{h}^\ast_{2-2}(f)\right) \\ 
    &= \frac{2 \pi c^3 r^2}{G} \int^{\infty}_{0} \diff f \:f^2 |\tilde{h}_{22}(f)|^2 \\
    &= \frac{2 \pi c^3 r^2}{G} \int^{\infty}_{0} \diff f \:f^2 {A^2_{\mathrm{GR}}(f)}\;.
\end{split}
\end{equation}
\end{widetext}

Similarly plugging Eqs.~\eqref{tilde_H} and \eqref{dot_tilde_H} into Eq.~~\eqref{dJz}, we derive
\begin{widetext}
\begin{equation}
\begin{split}
    \Delta J^z &= -\frac{c^3 r^2}{16\pi G} \: 
    \mathrm{Re} \left[ 
    \int^{\infty}_{-\infty} \diff t  \int \diff \Omega \: \frac{\partial {H}}{\partial \phi} \dot{{H}^\ast} \right] \\
    &= -\frac{c^3 r^2}{16\pi G} \: 
    \mathrm{Re} \left[ \int^{\infty}_{-\infty} \diff t  \int \diff \Omega \left[
    \begin{split}
    &\frac{\partial}{\partial \phi} \left(\int^{\infty}_{-\infty} \diff f \left({}^{-2}Y_{22}(\theta,\phi)\tilde{h}_{22}(f) + {}^{-2}Y_{2-2}(\theta,\phi)\tilde{h}_{2-2}(f) \right) e^{-2\pi i f t}\right) \\ &\cdot{} \left( \int^{\infty}_{-\infty} \diff f' \:(2\pi i f)\left({}^{-2}Y^\ast_{22}(\theta,\phi)\tilde{h}^\ast_{22}(f') -{}^{-2}Y^\ast_{2-2}(\theta,\phi)\tilde{h}^\ast_{2-2}(f') \right) e^{2\pi i f' t} \right) 
    \end{split}
    \right] \right]\\
    &=  \frac{c^3 r^2}{4 G} \: 
    \mathrm{Re} \left[ \iint^\infty_{-\infty} \diff f \diff f' \left( f'\tilde{h}_{22}(f)\tilde{h}^\ast_{22}(f') - f'\tilde{h}^\ast_{2-2}(f) \tilde{h}_{2-2}(f') \right) \int^{\infty}_{-\infty} \diff t \:e^{2\pi i (f-f')t} \right] \\
    &= \frac{c^3 r^2}{4 G} \: 
    \mathrm{Re} \left[ \int^\infty_{-\infty} \diff f \left( f\tilde{h}_{22}(f)\tilde{h}^\ast_{22}(f) - f\tilde{h}^\ast_{2-2}(f) \tilde{h}_{2-2}(f) \right) \right] \\
    &= \frac{2 c^3 r^2}{G} \int^\infty_{0} \diff f \: f|\tilde{h}_{22}(f)|^2  \\
    &= \frac{2 c^3 r^2}{G} \int^\infty_{0} \diff f \: fA^2_\mathrm{GR}(f)\;. \\
\end{split}
\end{equation}
\end{widetext}

\newpage


\nocite{*}

\newpage
\bibliographystyle{apsrev4-2}
\bibliography{reference}

\end{document}